\documentclass[pra,showpacs,twocolumn]{revtex4}
\usepackage{amsmath}
\usepackage{amssymb}
\usepackage{graphicx}
\usepackage{dcolumn}
\usepackage{bm}
\begin{document}
\title{Analytic, Group-Theoretic
Density Profiles for Confined, Correlated $N$-Body Systems}
\author{W.B.\ Laing, M.\ Dunn, D.K.\ Watson \\
University of Oklahoma \\
Homer L.\ Dodge Department of Physics and Astronomy \\
Norman, OK 73019}
\date{\today}

\begin{abstract}
Confined quantum systems involving $N$ identical interacting
particles are to be found in many areas of physics, including
condensed matter, atomic and chemical physics. A beyond-mean-field
perturbation method that is applicable, in principle, to weakly,
intermediate, and strongly-interacting systems has been set forth
by the authors in a previous series of papers. Dimensional
perturbation theory was used, and in conjunction with group
theory, an analytic beyond-mean-field correlated wave function at
lowest order for a system under spherical confinement with a
general two-body interaction was derived. In the present paper, we use
this analytic wave function to derive the corresponding
lowest-order, analytic density profile and apply it to the example
of a Bose-Einstein condensate.
\end{abstract}
\maketitle

\section{Introduction}
Confined quantum systems are widespread across many areas of
physics. They include, among other examples, quantum
dots\cite{quantumdots}, two-dimensional electronic systems in a
corbino disk geometry\cite{corbino}, Bose-Einstein condensates
(BECs)\cite{LegPit}, rotating superfluid helium
systems\cite{rsfhs}, and atoms (where the massive nucleus provides
the confining potential for the electrons).
These new (and
sometimes old) systems possess from a few tens to millions of
particles and present a challenge for existing $N$-body methods
when the mean-field approach\linebreak fails\cite{MinguzziAnderson}.

These systems span an enormous range in interparticle interaction
strength. For example, a gaseous BEC is typically a
weakly-interacting system for which a mean-field description is
perfectly adequate. On the other hand, a superfluid helium system
is a strongly interacting system.
In this regard we note that the
scattering length of a gaseous BEC
tuned by a
magnetic field can cover the entire range of interactions, from
weakly interacting to strongly interacting. A few of the methods
that have been applied to strongly-interacting systems
include the coupled cluster method (CCM)\cite{CC}, the method of
correlated basis functions (CBF)\cite{cbf}, density functional
theory\cite{singh}, and quantum Monte-Carlo
methods\cite{MinguzziAnderson,mc,blume}.

Dimensional perturbation
theory (DPT)\cite{copen92, chattrev} provides a systematic
approach to the study of correlation in quantum confined systems
This method
takes advantage of the high
degree of symmetry possible among identical particles in higher dimensions
to obtain an analytic description of the confined quantum system -
without making any assumptions about the number of particles or
the strength of interparticle interactions.  Because the
perturbation parameter is the inverse of the dimensionality of
space ($\delta = 1/D$), DPT is equally applicable to weakly or
strongly interacting systems.
Another important advantage of DPT is that
low-order DPT calculations are essentially {\em analytic}
in nature\cite{FGpaper}. As a consequence the number of atoms
enters into the calculations as a parameter, and so, in principle,
results for any $N$ are obtainable from a single
calculation\cite{energy}.
Also, even the lowest-order result
includes correlation, and so DPT may also be used to explore the
transition between weakly-interacting systems and those which are
strongly interacting.
These results can be systematically improved by going to higher
order\cite{matrix_method}.

In principle, excited states at low order in DPT are obtained
from the same analytic calculation\cite{energy}. They only differ
in the number of quanta in the different normal modes. Higher
orders in the DPT expansion of excited states can also be
calculated\cite{matrix_method}.

In past papers we have begun to apply dimensional perturbation
theory to calculate the first-order ground state energy and normal mode
frequencies of spherically-confined quantum
systems\cite{loeser,FGpaper,energy,paperI,prl1}.
In our last paper, we developed the formalism for deriving the lowest-order
DPT wave function for an $L = 0$ spherically confined quantum
system (see Ref.~\cite{paperI} here also referred to as Paper I).
In the present paper, we use the formalism developed in Paper I to
calculate the lowest-order DPT density profile of the spherically
confined quantum system.  We apply these results to the specific case of a
BEC for which the density profile is a directly observable manifestation
of the quantized behavior of the confined quantum system.

Section~\ref{sec:toolbox} provides a brief overview of the
tools used.  Section~\ref{sec:SE} discusses the
$L=0$, $D$-dimensional, $N$-body Schr\"odinger equation where $L$
is the total angular momentum of the system.
Sections~\ref{sec:infD}-\ref{sec:NormCalc} provide a summary of
the relevant results from our previous work, particularly
from Paper~I. Then in Section~\ref{sec:densityprofile}  the
lowest-order density profile is derived from the lowest-order wave function
of Paper~I (summarized Secs.~\ref{sec:infD}-\ref{sec:NormCalc}),
while Section~\ref{sec:optimize} discusses how it is possible to optimize
our lowest-order result by minimizing the contribution from
higher-order terms. Sections~\ref{sec:BECdensityprofile} and
\ref{sec:fitting} apply these results to the case of a BEC.
Section~\ref{sec:BECdensityprofile} sets up the problem for a BEC,
assuming a hard-sphere interatomic potential at $D=3$\,, while
Section~\ref{sec:fitting} discusses the details of minimizing
the contribution from higher-order terms.
Section~\ref{sec:resultsranddiscussion} discusses the results for
a BEC. Section~\ref{sec:Conc} presents our conclusions.

\section{Toolbox} \label{sec:toolbox}
The tools used to describe large-$N$ correlated wave functions are
carefully chosen to maximize the use of symmetry and minimize the
dependence on numerical computation. We handle the massive number
of interactions for large $N$ ($\sim N^2/2$ two-body interactions)
by bringing together three theoretical methods.

The first, DPT\cite{copen92, chattrev}, is chosen because its
$D\rightarrow\infty$ \linebreak equation yields a
maximally-symmetric configuration for $N$ identical particles
which allows an analytic solution. Higher orders yield insight
into fundamental motions as well as a framework for successive
approximations. The second method is the $FG$ method of Wilson,
Decius, and Cross\cite{dcw}. This seminal method has long been
used in quantum chemistry to study vibrations of polyatomic
molecules. It directly relates the structure of the Schr\"odinger
equation to the coordinate set which describes the normal modes of
the system. The third method, the use of group theoretic
techniques\cite{dcw,hamermesh}, takes full advantage of the
point-group symmetry in the $D\rightarrow\infty$ limit.

\section{The ${\bm{D}}$-dimensional ${\bm{N}}$-body Schr\"odinger
Equation} \label{sec:SE} The Schr\"odinger equation for an
$N$-body system of particles confined by a spherically symmetric
potential with a two-body potential $g_{ij}$\, is
\begin{equation}
\label{generalH} H \Psi = \left[ \sum\limits_{i=1}^{N} h_{i} +
\sum_{i=1}^{N-1}\sum\limits_{j=i+1}^{N} g_{ij} \right] \Psi = E
\Psi \,.
\end{equation}
In a $D$-dimensional Cartesian space
\begin{equation} \label{eq:hi}
h_{i}=-\frac{\hbar^2}{2
m_{i}}\sum\limits_{\nu=1}^{D}\frac{\partial^2}{\partial
x_{i\nu}^2} +
V_{\mathtt{conf}}\left(\sqrt{\sum\nolimits_{\nu=1}^{D}x_{i\nu}^2}\right)
\,,
\end{equation}
\begin{equation}
\mbox{and} \;\;\;
g_{ij}=V_{\mathtt{int}}\left(\sqrt{\sum\nolimits_{\nu=1}^{D}\left(x_{i\nu}-x_{j\nu}
\right)^2}\right)
\end{equation}
are the single-particle Hamiltonian and the two-body interaction
potential, respectively, $x_{i\nu}$ is the $\nu^{th}$ Cartesian component
of the $i^{th}$ particle, and $V_{\mathtt{conf}}$ is the confining
potential.

\subsection{The Effective $\bm{S}$-Wave Schr\"{o}dinger Equation.}
It is desirable to transform from Cartesian to internal coordinates.
A convenient internal coordinate system for confined
spherically symmetric ($L=0$) systems is
%
\begin{equation}\label{eq:int_coords}
\renewcommand{\arraystretch}{1.5}
\begin{array}{rcl} r_i & = &\sqrt{\sum_{\nu=1}^{D} x_{i\nu}^2}\,, \;\;\; (1 \le i \le
N)\,,
\;\;\; \mbox{and} 
\\ \gamma_{ij} & = & cos(\theta_{ij})=\left(\sum_{\nu=1}^{D}
x_{i\nu}x_{j\nu}\right) / r_i r_j\,,
\end{array}
\renewcommand{\arraystretch}{1}
\end{equation}
%
$(1 \le i < j \le N)$\,, which are the $D$-dimensional scalar
radii $r_i$ of the $N$ particles from the center of the confining
potential and the cosines $\gamma_{ij}$ of the $N(N-1)/2$ angles
between the radial vectors. Under this coordinate change the
effective $S$-wave Schr\"{o}dinger equation in these internal
coordinates becomes
\begin{widetext}
\begin{equation} \label{eq:laplacian} \begin{array}{r@{}l} {\displaystyle \left[ \sum\limits_{i} \left\{ -\frac{\hbar^2}{2
m_{i}} \left( \vphantom{\frac{D-1}{r_i^2}\sum\limits_{j\not=
i}\gamma_{ij}\frac{\partial} {\partial \gamma_{ij}}} \right.
\right. \right. } & {\displaystyle \frac{\partial^2}{{\partial
r_i}^2} + \frac{D-1}{r_i}\frac{\partial}{\partial r_i} +
\sum\limits_{j\not= i}\sum\limits_{k\not=
i}\frac{\gamma_{jk}-\gamma_{ij}\gamma_{ik}}
{r_i^2}\frac{\partial^2}{\partial \gamma_{ij} \partial
\gamma_{ik}} } \\
&  {\displaystyle \left. \left. \left. -
\frac{D-1}{r_i^2}\sum\limits_{j\not= i}\gamma_{ij}\frac{\partial}
{\partial \gamma_{ij}} \right) + V_{\mathtt{conf}} (r_i) \right\}
+ \sum_{i=1}^{N-1}\sum\limits_{j=i+1}^{N} V_{\mathtt{int}}(r_{ij})
\right] \Psi = E \Psi \,,}
\end{array}
\end{equation}
\end{widetext}
where $(r_{ij})^2 = (r_i)^2 + (r_j)^2 - \,\, 2 \, r_i \, r_j \,
\gamma_{ij}$\,.

\subsection{The Jacobian-Weighted Schr\"{o}dinger Equation}
\label{sub:simtransf} Dimensional perturbation theory utilizes a
similarity transformation so that the kinetic energy operator is
transformed into a sum of two types of terms, namely, derivative
terms {\em and} a repulsive centrifugal-like term. The latter
repulsive centrifugal-like term stabilizes the system against
collapse in the large-$D$ limit when attractive interparticle
potentials are present. Low orders of the dimensional ($1/D$)
expansion of the similarity-transformed Schr\"{o}dinger equation
are then exactly soluble for any value of $N$\,. In the
$D\to\infty$ limit the derivative terms drop out resulting in a
static problem, while large-$D$ corrections correspond to simple
harmonic normal-mode oscillations about the
infinite-dimensional\pagebreak[4]
structure. (See Sections IV - VII below.)

In Paper~I the weight function was chosen to be the square root of
the inverse of the Jacobian, $J$\,, where\cite{avery} $J = (r_1
r_2 \ldots r_N)^{D-1} \Gamma^{(D-N-1)/2}$ and $\Gamma$ is the
Gramian determinant, the determinant of the matrix whose elements
are $\gamma_{ij}$ (see Appendix D of Ref.~\cite{FGpaper}), so that
the similarity-transformed wave function $\Phi$ and operators
$\widetilde{O}$ are $\Phi = J^{\frac{1}{2}} \, \Psi \;\;
\mbox{and} \;\; \widetilde{O}= J^{\frac{1}{2}} \, \widehat{O} \,
J^{-\frac{1}{2}}$\,, respectively. Under this Jacobian transformation, a first
derivative of an internal coordinate is the conjugate momentum to
that coordinate. The matrix elements of coordinates and their
derivatives between the lowest-order normal-mode functions, which
are involved in the development of higher-order DPT expansions,
are much easier to calculate since the weight function in the
integrals is now unity.

Carrying out the above Jacobian transformation of the
Schr\"{o}dinger equation of Eq.~(\ref{generalH}), we
obtain\cite{avery}:
\begin{equation}
 (T+V)\, \Phi = E \,\, \Phi
\label{eq:SE}
\end{equation}
where
\begin{widetext}
\begin{equation}
T 
= 
{\displaystyle \hbar^2
\sum\limits_{i=1}^{N}\Biggl[-\frac{1}{2
m_i}\frac{\partial^2}{{\partial r_i}^2}- \frac{1}{2 m_i r_i^2}
\sum\limits_{j\not=i}\sum\limits_{k\not=i}
\frac{\partial}{\partial\gamma_{ij}}(\gamma_{jk}-\gamma_{ij}
\gamma_{ik})
\frac{\partial}{\partial\gamma_{ik}}}
{\displaystyle +\frac{N(N-2)+(D-N-1)^2 \left(
\frac{\Gamma^{(i)}}{\Gamma} \right) }{8 m_i r_i^2} \Biggr] }
 \label{eq:SE_T}
\end{equation}
\end{widetext}
and
\begin{equation}
V=\sum\limits_{i=1}^{N}V_{\mathtt{conf}}(r_i)+
\sum\limits_{i=1}^{N-1}\sum\limits_{j=i+1}^{N}
V_{\mathtt{int}}(r_{ij}) \,.
\end{equation}
Equation~(\ref{eq:SE_T}) for $T$ is explicitly self-adjoint
since the weight function, $W$\,, for the matrix elements is equal
to unity. The similarity-transformed Hamiltonian for the energy
eigenstate $\Phi$ is $ H=( T+V)$\,.

\section{Infinite-${\bm{D}}$ analysis: Leading order
energy} \label{sec:infD}

Following Paper I, we begin the perturbation analysis by
defining dimensionally scaled variables:

\begin{equation} \label{eq:kappascale}
\bar{r}_i = r_i/\kappa(D) \,, \;\;\; \bar{E} = \kappa(D) E \,,
\;\;\; \mbox{and} \;\;\; \bar{H} = \kappa(D) \,\, H
\end{equation}
where $\kappa(D)$ is a dimension-dependent scale factor, which
regularizes the large-dimension limit.
From
Eq.~(\ref{eq:SE_T}) the kinetic energy T scales in the same way as
$1/r^2$\,, so the scaled version of Eq.~(\ref{eq:SE}) becomes
\begin{equation} \label{eq:scale1}
\bar{H} \Phi =
\left(\frac{1}{\kappa(D)}\bar{T}+\bar{V}_{\mathtt{eff}}
\right)\Phi = \bar{E} \Phi,
\end{equation}
where barred quantities indicate that the variables are now in
scaled units. The centrifugal-like term in $T$ of
Eq.~(\ref{eq:SE_T}) has
 quadratic $D$ dependence so the scale factor $\kappa(D)$
 must also be quadratic in $D$\,,
otherwise the $D\to\infty$ limit of the Hamiltonian would not be
finite. The precise form of $\kappa(D)$ depends on the particular system
and is chosen so that the
result of the scaling is as simple as possible.
In previous work\cite{FGpaper} we have chosen
$\kappa(D)=(D-1)(D-2N-1)/(4Z)$ for the $S$-wave, $N$-electron
atom; $\Omega \, l_{\mathtt{ho}}$ for the $N$-electron quantum dot
where $\Omega=(D-1)(D-2N-1)/4$ and the dimensionally-scaled
harmonic oscillator length and trap frequency respectively are
${l}_{\mathtt{ho}}=\sqrt{\frac{\hbar}{m^*\bar{\omega}_{\mathtt{ho}}}}$
and $\bar{\omega}^2_{\mathtt{ho}}=\Omega^3
{\omega}^2_{\mathtt{ho}}$\,; and $D^2 \bar{a}_{\mathtt{ho}}$ for
the BEC where $\bar{a}_{\mathtt{ho}}=\sqrt{\frac{\hbar}{m
\bar{\omega}_{\mathtt{ho}}}}$ and
${\bar{\omega}_{\mathtt{ho}}}=D^3{\omega_{\mathtt{ho}}}$\,. In the
$\delta \rightarrow 0$ ($D \rightarrow \infty$) limit, where
\begin{equation}
\label{delta}
\delta \equiv 1/D \,,
\end{equation}
the factor of $\kappa(D)$ in the denominator of Eq.~(\ref{eq:scale1})
suppresses the derivative terms leaving behind a
centrifugal-like term in an effective potential,
\begin{equation}
\label{veff}
\renewcommand{\arraystretch}{1.5}
\begin{array}[b]{r@{\hspace{0.4ex}}c@{\hspace{0.2ex}}l}
{\displaystyle \bar{V}_{\mathtt{eff}}(\bar{r},\gamma;\delta=0)}&=&
{\displaystyle \sum\limits_{i=1}^{N}\left(\frac{\hbar^2}{8 m_i
\bar{r}_i^2}\frac{\Gamma^{(i)}}{\Gamma}+\bar{V}_{\mathtt{conf}}(\bar{r},\gamma;\delta=0)\right)
}
\\ && {\displaystyle +\sum\limits_{i=1}^{N-1}\sum\limits_{j=i+1}^{N}
\bar{V}_{\mathtt{int}}(\bar{r},\gamma;\delta=0)\,, }
\end{array} \renewcommand{\arraystretch}{1}
\end{equation}
in which the particles become frozen.
In the $D\to\infty$ ($\delta \to 0$) limit, the excited
states collapse onto the ground state at the minimum of
$V_{\mathtt{eff}}$\,.

We assume a totally symmetric, large-dimension configuration
at which the effective potential is a minimum.  The $N$
particles are arranged on a hypersphere, each particle with a radius,
$\bar{r}_{\infty}$, from the center of the confining potential.
Furthermore, the angle cosines between each pair of particles takes
on the same value, $\overline{\gamma}_{\infty}$, i.e.\
\begin{equation} \label{eq:Lewis}
 \renewcommand{\arraystretch}{1.5}
\begin{array}{rcl}
{\displaystyle \lim_{D \rightarrow \infty} \bar{r}_{i} } & = &
{\displaystyle \bar{r}_{\infty} \;\; (1 \le i \le N)\,, } \\
{\displaystyle \lim_{D \rightarrow \infty} \gamma_{ij} } & = &
{\displaystyle \overline{\gamma}_{\infty} \;\; (1 \le i < j \le
N)\,. }
\end{array}
 \renewcommand{\arraystretch}{1}
\end{equation}
This configuration for $N=1,$ $2,$ $3,$ and $4$ is comparable to
the sequence of arrangements of hydrogen atoms in LiH, H$_2$O,
NH$_3$\,, and CH$_4$\,. In this analogy, the heterogeneous atom
represents the center of the confining potential\cite{footn}.
(This symmetric high-dimensional structure is also not unlike the
localized structure found in a hyperspherical treatment of the
confined two-component normal Fermi gas in the $N \rightarrow
\infty$ limit\cite{seth}.)

In scaled units the $D\to\infty$ approximation for the energy is
simply the effective potential minimum, i.e.
\begin{equation}
\label{zeroth}
\bar{E}_{\infty}=\bar{V}_{\mathtt{eff}}(\bar{r}_{\infty},\overline{\gamma}_{\infty};
\hspace{1ex} \delta=0)\,.
\end{equation}
In this $D\rightarrow\infty$ approximation, the centrifugal-like
term that appears in $\bar{V}_{\mathtt{eff}}$\,, which is nonzero even for the
ground state, is a zero-point energy contribution satisfying the
minimum uncertainty principle\cite{chat}.

Beyond-mean-field effects are already present in this
approximation. This may be seen in the value of
$\overline{\gamma}_{\infty}$\,, the $D\rightarrow\infty$
expectation value for the interparticle angle cosine (see
Eqs.~(\ref{eq:Lewis}) and (\ref{eq:taylor1})). In the mean-field
approximation the expectation value for the interparticle angle
cosine for the $L=0$ states considered in this paper is zero. The
fact that $\overline{\gamma}_{\infty}$ is {\em not} zero is an
indication that beyond-mean-field effects are included even in the
$D\rightarrow\infty$ limit.

This symmetric structure in which all $N$ particles are
equidistant and equiangular from every other particle can only
exist in a higher-dimensional space and is impossible in a
three-dimensional space unless $N \leq 4$\, (see above). That this
high dimensional structure makes sense can be seen as follows.
As we have noted above, $\bar{r}_{\infty}$ and
$\overline{\gamma}_{\infty}$ are the lowest-order DPT expectation
values for $\bar{r}_i$ and $\gamma_{ij}$\,. This would indicate
that the expectation values for the radii and interparticle angles
for actual $D=3$ systems should have values corresponding to structures
that can only exist in higher ($D > 3$) dimensional spaces. Accurate
configuration interaction calculations for atoms in three dimensions
do indeed have expectation
values for the radii and inter-electron angles 
which define higher dimension structures\cite{higherD}.

This highly-symmetric, $D \rightarrow \infty$ structure imparts a
point-group structure to the system which is isomorphic to the
symmetric group of $N$ identical objects\cite{hamermesh}, $S_N$, and allows
for a largely analytic solution to the problem, even though the
number of degrees of freedom becomes very large when $N$ is large.
The $D\rightarrow\infty$ approximation may also be systematically
improved by using it as the starting point for a perturbation
expansion (DPT)\cite{matrix_method}. In this regard, the $S_N$
symmetry greatly simplifies this task since the interaction terms
individually have to transform as a scalar under the $S_N$ point
group.

\section{Normal-mode analysis and the ${\bm{1/D}}$ next-order energy
correction}\label{sec:firstorder}

In the $D \rightarrow \infty$ limit, the particles are frozen in a
completely symmetric configuration (which is somewhat analogous
to the Lewis structure in chemical terminology\cite{lewis}). This
configuration determines the $D \rightarrow \infty$ energy,\,
$\bar{E}_{\infty}$\,, but it is not a wave function as no nodal
information is present. The large-dimension limit is a singular
limit of the theory, so to obtain the lowest-order wave function
we have to consider the next order in the perturbation expansion,
which yields not only the $(1/D)^0$\,, i.e.\ leading-order wave
function, but the order $(1/D)^1$ correction to the energy.
The perturbation series has the form:
\begin{equation} \renewcommand{\arraystretch}{2} \begin{array}{r@{}l@{}c}
{\displaystyle \bar{E} = \bar{E}_{\infty} + \delta} &
{\displaystyle \, \sum_{j=0}^\infty
\left(\delta^{\frac{1}{2}}\right)^j \,
\bar{E}_j } & \\
{\displaystyle \Phi(\bar{r}_i,\gamma_{ij}) = } & {\displaystyle \,
\sum_{j=0}^\infty \left(\delta^{\frac{1}{2}}\right)^j \, \Phi_j }& \,.
\end{array}
\renewcommand{\arraystretch}{1}
\end{equation}
In practice $\bar{E}_j=0$ $\forall$ $j$ odd. This next-order $1/D$
correction in the energy, but lowest-order in the wave function,
can be viewed as involving small oscillations about the $D
\rightarrow \infty$ structure (somewhat analogous to Langmuir
oscillations\cite{langmuir}).
As we shall see below, these are obtained from a harmonic equation, and so
we refer to them as the energy and wave function at harmonic
order.

To obtain this harmonic correction to the energy for large but
finite values of $D$\,, we expand about the minimum of the
$D\to\infty$ effective potential. We first define a position
vector, ${\bar{\bm{y}}}$\,, consisting of all $N(N+1)/2$ internal
coordinates:
\begin{equation}\label{eq:ytranspose}
\begin{array}[t]{l} {\bar{\bm{y}}} = \left( \begin{array}{c} \bar{\bm{r}} \\
\bm{\gamma} \end{array} \right) \,, \;\;\; \mbox{where} \;\;\; \\
\mbox{and} \;\;\; \bar{\bm{r}} = \left(
\begin{array}{c}
\bar{r}_1 \\
\bar{r}_2 \\
\vdots \\
\bar{r}_N
\end{array}
\right) \,. \end{array} 
\bm{\gamma} =
\left(
\begin{array}{c}
\gamma_{12} \\ \cline{1-1}
\gamma_{13} \\
\gamma_{23} \\ \cline{1-1}
\gamma_{14} \\
\gamma_{24} \\
\gamma_{34} \\ \cline{1-1}
\gamma_{15} \\
\gamma_{25} \\
\vdots \\
\gamma_{N-2,N} \\
\gamma_{N-1,N} \end{array} \right) \,,
\end{equation}
We then make the following substitutions for all radii and angle
cosines:
\begin{equation} \label{eq:taylor1}
\bar{r}_{i} = \bar{r}_{\infty}+\delta^{1/2}\bar{r}'_{i} \;\;\;
\mbox{and} \;\;\; \gamma_{ij} =
\overline{\gamma}_{\infty}+\delta^{1/2}\overline{\gamma}'_{ij} \,.
\end{equation}
In all practical
situations $\bar{V}_{\mathtt{eff}}({\bar{\bm{y}}}; \delta)$ is a
function of ${\bar{\bm{y}}}$ and $\delta$, so we may obtain a
power series in $\delta^{1/2}$ of the effective potential about
the $D\to\infty$ symmetric minimum.

Defining a displacement vector of the internal displacement
coordinates [primed in Eqs. (\ref{eq:taylor1})]
\begin{equation}\label{eq:ytransposeP}
\begin{array}[t]{l} {\bar{\bm{y}}'} = \left( \begin{array}{c} \bar{\mathbf{r}}' \\
\overline{\bm{\gamma}}' \end{array} \right) \,, \;\;\;
\mbox{where} \;\;\; \\ \mbox{and} \;\;\; \bar{\bm{r}}' = \left(
\begin{array}{c}
\bar{r}'_1 \\
\bar{r}'_2 \\
\vdots \\
\bar{r}'_N
\end{array}
\right) \,, \end{array} 
\overline{\bm{\gamma}}' = \left(
\begin{array}{c}
\overline{\gamma}'_{12} \\ \cline{1-1}
\overline{\gamma}'_{13} \\
\overline{\gamma}'_{23} \\ \cline{1-1}
\overline{\gamma}'_{14} \\
\overline{\gamma}'_{24} \\
\overline{\gamma}'_{34} \\ \cline{1-1}
\overline{\gamma}'_{15} \\
\overline{\gamma}'_{25} \\
\vdots \\
\overline{\gamma}'_{N-2,N} \\
\overline{\gamma}'_{N-1,N} \end{array} \right) \,,
\end{equation}
we find
\begin{eqnarray}
\lefteqn{\bar{V}_{\mathtt{eff}}({\bar{\bm{y}}'}; \hspace{1ex}
\delta) = \left[ \bar{V}_{\mathtt{eff}} \right]_{\delta^{1/2}=0} }
\nonumber \\ && + \frac{1}{2} \, \delta \left\{
\sum\limits_{\mu=1}^{P} \sum\limits_{\nu=1}^{P} \bar{y}'_{\mu}
\left[\frac{\partial^2 \bar{V}_{\mathtt{eff}}}{\partial
\bar{y}_{\mu}
\partial \bar{y}_{\nu}}\right]_{\delta^{1/2}=0} \hspace{-1.5em} \bar{y}'_{\nu} + v_o \right\} +
O\left(\delta^{3/2}\right) \,, \nonumber \\ \label{Taylor}
\end{eqnarray}
where $P \equiv N(N+1)/2$ is the number of internal coordinates.
The quantity, $v_o$, is just a constant term, independent of
$\bar{y}'_{\mu}$, while
the first term of the $O((\delta^{1/2})^2)$ term defines the
elements of the Hessian matrix\cite{strang} $\bm{F}$ of
Eq.~(\ref{Gham}) below. The derivative terms in the kinetic energy are
taken into account by a similar series expansion, beginning with a
harmonic-order term that is bilinear in ${\partial/\partial
\bar{y}'}$\,, i.e.\
\begin{equation}\label{eq:T}
{\mathcal T}=-\frac{1}{2} \delta \sum\limits_{\mu=1}^{P}
\sum\limits_{\nu=1}^{P} {G}_{\mu\nu}
\partial_{\bar{y}'_{\mu}}
\partial_{\bar{y}'_{\nu}} + O\left(\delta^{3/2}\right),
\end{equation}
where ${\mathcal T}$ is the derivative portion of the kinetic
energy $T$ {see Eq.~(\ref{eq:SE_T})). It follows from Eqs.
(\ref{Taylor}) and (\ref{eq:T}) that $\bm{G}$ and $\bm{F}$\,, both
constant matrices, are defined in the harmonic-order Hamiltonian
as follows:
\begin{equation}\label{Gham}
\widehat{H}_1=-\frac{1}{2} {\partial_{\bar{y}'}}^{T} \bm{G}
{\partial_{\bar{y}'}} + \frac{1}{2} \bar{\bm{y}}^{\prime T} {\bm
F} {{\bar{\bm{y}}'}} + v_o \,.
\end{equation}
Thus, obtaining the harmonic-order wave function, and
energy correction is reduced to
solving a harmonic equation by finding the normal modes of the
system.

\subsection{The FG Matrix Method for the Normal Modes and Frequencies}
We use the FG matrix method\cite{dcw} to obtain the normal-mode
vibrations and, thereby, the harmonic-order energy correction. A
review of the FG matrix method is presented in Appendix A of Paper~I,
but the main results may be stated as follows. The $b^{\rm th}$ normal
mode coordinate may be written as (Eq.~(A9) of Paper~I)
\begin{equation} \label{eq:qyt}
[{\bm q'}]_b = {\bm{b}}^T {\bar{\bm{y}}'} \,,
\end{equation}
where the coefficient vector ${\bm{b}}$ satisfies the
eigenvalue equation (Eq.~(A10) of Paper~I)
\begin{equation} \label{eq:FGit}
{\bf F} \, \bm{G} \, {\bm{b}} = \lambda_b \, {\bm{b}}
\end{equation}
with the resultant secular equation (Eq.~(A11) of Paper~I)
$\det({\bf F}\bm{G}-\lambda{\bf I})=0$\,. The coefficient vector
also satisfies the normalization condition (Eq.~(A12) of Paper~I)
${\bm{b}}^T \bm{G} \, {\bm{b}} = 1$\,. As can be seen from
Eq.~(A3) of Paper~I the frequencies are given by
$\lambda_b=\bar{\omega}_b^2$\,.

In an earlier paper\cite{FGpaper} we solve these equations for the
frequencies. The number of roots $\lambda$ of the secular equation
-- there are $P \equiv N(N+1)/2$ roots -- is potentially huge.
However, due to the $S_N$ symmetry of the problem discussed in
Sect.~5 and 6 of Paper~I, there is a very significant
simplification. The secular equation has only five distinct roots,
$\lambda_{\mu}$\,, where $\mu$ is a label which
runs over ${\bf 0}^-$\,, ${\bf
0}^+$\,, ${\bf 1}^-$\,, ${\bf 1}^+$\,, and ${\bf 2}$\,, regardless
of the number of particles in the system (see Refs.~\cite{FGpaper}
and Sect.~\ref{subsec:symnorm}). Thus the energy through
harmonic-order (see Eq.~(\ref{eq:E1})) can be written in terms of
the five distinct normal-mode vibrational frequencies which are
related to the roots $\lambda_{\mu}$ of $FG$ by
\begin{equation}\label{eq:omega_p}
\lambda_{\mu}=\bar{\omega}_{\mu}^2 \,.
\end{equation}
The energy through harmonic order in $\delta$ is
then\cite{FGpaper}
\begin{equation}
\overline{E} = \overline{E}_{\infty} + \delta \Biggl[
\sum_{\renewcommand{\arraystretch}{0}
\begin{array}[t]{r@{}l@{}c@{}l@{}l} \scriptstyle \mu = \{
  & \scriptstyle \bm{0}^\pm,\hspace{0.5ex}
  & \scriptstyle \bm{1}^\pm & , & \\
  & & \scriptstyle \bm{2} & & \scriptstyle  \}
            \end{array}
            \renewcommand{\arraystretch}{1} }
\hspace{-0.50em} \sum_{\mathsf{n}_{\mu}=0}^\infty
({\mathsf{n}}_{\mu}+\frac{1}{2}) d_{\mu,\mathsf{n}_{\mu}}
\bar{\omega}_{\mu} \, + \, v_o \Biggr] \,, \label{eq:E1}
\end{equation}
where the $\mathsf{n}_{\mu}$ are the vibrational quantum numbers
of the normal modes of the same frequency $\bar{\omega}_{\mu}$
($\mathsf{n}_{\mu}$ counts the number of nodes in a given normal
mode). The quantity $d_{\mu,\mathsf{n}_{\mu}}$ is the occupancy of
the manifold of normal modes with vibrational quantum number
$\mathsf{n}_{\mu}$ and normal mode frequency $\bar{\omega}_{\mu}$,
i.e.\ it is the number of normal modes with the same frequency
$\bar{\omega}_{\mu}$ and the same number of quanta
$\mathsf{n}_{\mu}$.  The total occupancy of the normal modes with
frequency $\bar{\omega}_{\mu}$ is equal to the multiplicity of the
root $\lambda_{\mu}$, i.e.\ $d_{\mu} =
\sum_{\mathsf{n}_{\mu}=0}^\infty d_{\mu,\mathsf{n}_{\mu}}$\,,
where $d_{\mu}$ is the multiplicity of the $\mu^{th}$ root. The
multiplicities of the five roots are\cite{FGpaper}
\begin{equation} \label{eq:dpm}
\renewcommand{\arraystretch}{1.5}
\begin{array}{l}
d_{{\bf 0}^+} = 1 \,, \hspace{1ex} d_{{\bf 0}^-} = 1 \,,
\hspace{1ex} d_{{\bf 1}^+} = N-1 \,, \\ d_{{\bf 1}^-} = N-1 \,,
\hspace{1ex} d_{{\bf 2}} = N(N-3)/2 \,. \end{array}
\renewcommand{\arraystretch}{1}
\end{equation}

An analysis of the character of the normal modes reveals that
the ${\bf 2}$ normal modes are phonon, i.e. compressional, modes;
the ${\bf 1^\pm}$ modes show single-particle character, and the
${\bf 0^\pm}$ normal modes in a harmonic confining potential
describe center-of-mass and breathing motions.\cite{excitations}

\section{The Symmetry of the $\bm{F}$ and $\bm{G}$ Matrices
  and the Jacobian-Weighted Wave Function}
\label{sec:symm}
\subsection{$\bm{Q}$ matrices in terms of simple invariant submatrices}
\label{subsec:Qsubm} Such a high degree of degeneracy of the
frequencies in large-$D$\,, $N$-body quantum confinement problems
indicates a very high degree of symmetry. The $\bm{F}$\,, $\bm{G}$\,,
and $\bm{FG}$ matrices, which we generically denote by $\bm{Q}$\,,
are $P \times P$ matrices. The $S_N$ symmetry of the $\bm{Q}$
matrices ($\bm{F}$\,, $\bm{G}$\,, and $\bm{FG}$) allows us to
write these matrices in terms of six simple submatrices which are
invariant under $S_N$\,. We first define the number of
$\gamma_{ij}$ coordinates to be $M \equiv N(N-1)/2$\,, and let
$\bm{I}_N$ be an $N \times N$ identity matrix, $\bm{I}_M$ an $M
\times M$ identity matrix, $\bm{J}_N$ an $N \times N$ matrix of
ones and $\bm{J}_M$ an $M \times M$ matrix of ones. Further, we
let $\bm{R}$ be an $N \times M$ matrix such that
${R}_{i,jk}=\delta_{ij}+\delta_{ik}$\,, $\bm{J}_{NM}$ be an $N
\times M$ matrix of ones, and $\bm{J}^T_{NM}=\bm{J}_{MN}$\,. These
matrices are invariant under interchange of the particles, effected
by the point group $S_N$. They also have a specific interpretation in the
context of spectral graph theory (see Appendix B of
Ref.~\cite{FGpaper}).

We can then write the $\bm{Q}$ matrices as
\begin{equation}\label{eq:Q}
{\bf Q}=\left(\begin{array}{cc} {\bf Q}_{\bar{\bm{r}}'
\bar{\bm{r}}'} & {\bf Q}_{\bar{\bm{r}}' \overline{\bm{\gamma}}'}
\\ {\bf Q}_{\overline{\bm{\gamma}}' \bar{\bm{r}}'} & {\bf
Q}_{\overline{\bm{\gamma}}' \overline{\bm{\gamma}}'}
\end{array}\right) = \left(\begin{array}{c|c} Q_{i,j}
 & Q_{i,jk}
\\ \hline Q_{ij,k} & Q_{ij,kl} \end{array}
\right)
\end{equation}
where the block $\bm{Q}_{\bar{\bm{r}}' \bar{\bm{r}}'}$ has
dimension $(N \times N)$\,, block $\bm{Q}_{\bar{\bm{r}}'
\overline{\bm{\gamma}}'}$ has dimension $(N \times M)$\,, block
$\bm{Q}_{\overline{\bm{\gamma}}' \bar{\bm{r}}'}$ has dimension $(M
\times N)$\,, and block $\bm{Q}_{\overline{\bm{\gamma}}'
\overline{\bm{\gamma}}'}$ has dimension $(M \times M)$\,. As shown
in Appendix~B of Ref.~\cite{FGpaper},
%
%
\begin{eqnarray}
{\bf Q}_{\bar{\bm{r}}'
\bar{\bm{r}}'} & = & (Q_a-Q_b) {\bf I}_N + Q_b {\bf J}_N \,, \label{eq:Qrr} \\
\lefteqn{ \left. \begin{array}{rcl} {\displaystyle {\bf
Q}_{\bar{\bm{r}}' \overline{\bm{\gamma}}'} }& =
& {\displaystyle (Q_e-Q_f) {\bf R} + Q_f {\bf J}_{NM} } \\
{\displaystyle {\bf Q}_{\overline{\bm{\gamma}}' \bar{\bm{r}}'} } &
= & {\displaystyle (Q_c-Q_d) {\bf R}^T + Q_d {\bf J}_{NM}^T }
\end{array} \right\} \,,} \hspace{2.85em}
\label{eq:Qgr} \\
{\bf Q}_{\overline{\bm{\gamma}}' \overline{\bm{\gamma}}'} & = &
(Q_g-2Q_h+Q_{\iota}) {\bf I}_M + (Q_h-Q_{\iota}) {\bf R}^T {\bf R}
\nonumber \\ && + \, Q_{\iota} {\bf J}_M \,. \label{eq:Qgg}
\end{eqnarray}
It is this structure that causes the remarkable reduction from
a possible $P=N(N+1)/2$ distinct frequencies to just five distinct
frequencies for $L = 0$ systems\cite{multipole}.

In particular, letting ${\bf Q}$ be $\bm{FG}$\,, the matrix to be
diagonalized, or $\bm{G}$\,, the matrix required for the
normalization condition, Eq.~(\ref{eq:Q}) becomes
\begin{equation} \label{eq:GFsub}
\hspace{-0.1ex} \begin{array}{l}  \bm{FG}= \left(
\begin{array}{cc}
\tilde{a} {\bf I}_N + \tilde{b} {\bf J}_N & \tilde{e} {\bf R} + \tilde{f} {\bf J}_{NM} \\
\tilde{c} {\bf R}^T + \tilde{d} {\bf J}_{MN} & \tilde{g} {\bf I}_M
+ \tilde{h} {\bf R}^T {\bf R} + \tilde{\iota} {\bf J}_M
\end{array}\right) \,, \\  \\ \mbox{or} \;\;\; \bm{G} = \left( \begin{array}{cc}
\tilde{a}' {\bf I}_N & \bm{0} \\
\bm{0} & \tilde{g}' {\bf I}_M + \tilde{h}' {\bf R}^T {\bf R}
\end{array} \right) \,, \end{array}
\end{equation}
where the coefficients
$\tilde{a}$\,, $\tilde{b}$\,, $\tilde{c}$\,, $\tilde{d}$\,,
$\tilde{e}$\,, $\tilde{f}$\,, $\tilde{g}$\,, $\tilde{h}$\,, and
$\tilde{\iota}$ have a simple relation to the elements of the
$\bm{F}$ and $\bm{G}$ matrices while $\tilde{a}'$\,,
$\tilde{g}'$\,, and $\tilde{h}'$ have a simple relation to the
elements of the $\bm{G}$ matrix (see Paper I). These coefficients
also depend on
the choice of $\kappa(D)$ (See Ref.~\cite{FGpaper}).

\subsection{Symmetry Coordinates, Normal Coordinates, and the Jacobian-Weighted Wave function}
\label{subsec:symnorm} As
discussed in Paper I the $\bm{FG}$ matrix is a $N(N+1)/2 \times
N(N+1)/2$ dimensional matrix (there being $N(N+1)/2$ internal
coordinates), and so the secular equation could have up to
$N(N+1)/2$ distinct frequencies. However, as noted above, there
are only five distinct frequencies. The $S_N$ symmetry is
responsible for the remarkable reduction from $N(N+1)/2$ possible
distinct frequencies to five actual distinct frequencies. As we
shall also see, the $S_N$ symmetry greatly simplifies the
determination of the normal coordinates and hence the solution of
the large-$D$ problem.

The $\bm{Q}$ matrices, and in particular the $\bm{FG}$ matrix, are
invariants under $S_N$\,, so they do not connect subspaces
belonging to different irreducible representations of
$S_N$\cite{WDC}. Thus from Eqs.~(\ref{eq:qyt}) and (\ref{eq:FGit})
the normal coordinates must transform under irreducible
representations of $S_N$\,. Since the normal coordinates will be
linear combinations of the elements of the internal coordinate
displacement vectors $\bar{\bm{r}}'$ and
$\overline{\bm{\gamma}}'$\,, we first look at the $S_N$
transformation properties of the internal coordinates.

The internal coordinate displacement vectors $\bar{\bm{r}}'$ and
$\overline{\bm{\gamma}}'$ of Eqs.~(\ref{eq:taylor1}) are basis
functions which transform under matrix representations of $S_N$\,,
and each span the corresponding carrier spaces, however these
representations of $S_N$  are not irreducible representations  of
$S_N$\,.

In Sec.~6.1 of Paper~I we have shown that the reducible
representation under which $\bar{\bm{r}}'$ transforms is reducible
to one $1$-dimensional irreducible representation labelled by the
partition $[N]$ (the partition denotes a corresponding Young
diagram ( = Young pattern = Young shape)) of an irreducible
representation (see Appendix~C of Paper~I) and one
$(N-1)$-dimensional irreducible representation labelled by the
partition $[N-1, \hspace{1ex} 1]$\,. We will also show that the
reducible representation under which $\overline{\bm{\gamma}}'$
transforms is reducible to one $1$-dimensional irreducible
representation labelled by the partition $[N]$\,, one
$(N-1)$-dimensional irreducible representation labelled by the
partition $[N-1, \hspace{1ex} 1]$\,, and one $N(N-3)/2$-dimensional
irreducible representation labelled by the partition $[N-2,
\hspace{1ex} 2]$\,. Thus if $d_\alpha$ is the dimensionality of
the irreducible representation of $S_N$ denoted by the partition
$\alpha$ then $d_{[N]} = 1$\,, \; $d_{[N-1, \hspace{1ex} 1]} =
 N-1$\,, \; and $d_{[N-2, \hspace{1ex} 2]} =
 N(N-3)/2$\,. We note that:
$d_{[N]} + d_{[N-1, \hspace{1ex} 1]} = N \,,$ giving correctly the
dimension of the $\bar{\bm{r}}'$ vector and that: $d_{[N]} +
d_{[N-1, \hspace{1ex} 1]} + d_{[N-2, \hspace{1ex} 2]} =
N(N-1)/2$\,, giving correctly the dimension of the
$\overline{\bm{\gamma}}'$ vector.

Since the normal modes transform under irreducible representations
of $S_N$ and are composed of linear combinations of the elements
of the internal coordinate displacement vectors $\bar{\bm{r}}'$
and $\overline{\bm{\gamma}}'$\,, there will be two $1$-dimensional
irreducible representations labelled by the partition $[N]$\,, two
$(N-1)$-dimensional irreducible representations labelled by the
partition $[N-1, \hspace{1ex} 1]$\,, and one entirely angular
$N(N-3)/2$-dimensional irreducible representation labelled by the
partition $[N-2, \hspace{1ex} 2]$\,. Thus if we look at
Eq.~(\ref{eq:dpm}) we see that the ${\bf 0}^\pm$ normal modes
transform under two $[N]$ irreducible representations, the ${\bf
1}^\pm$ normal modes transform under two $[N-1, \hspace{1ex} 1]$
irreducible representations, while the ${\bf 2}$ normal modes
transform under the $[N-2, \hspace{1ex} 2]$ irreducible
representation.

The wave function for the harmonic-order Hamiltonian of
Eq.~(\ref{Gham}) is thus the product of $P=N(N+1)/2$
harmonic-oscillator wave functions:
\begin{equation} \label{eq:wavefunct}
\Phi_0({\bar{\bm{y}}}^{\prime})= \prod_{ \mu = \{
\bm{0}^\pm,\hspace{0.5ex} \bm{1}^\pm,\hspace{0.5ex} \bm{2} \} }
 \,\, \prod_{\xi=1}^{d_{\mu}} \,\, \phi_{n_{\mu_\xi}}
\!\! \left( \sqrt{\bar{\omega}_\mu} \,\,\, [{\bm{q}'}^\mu]_\xi
\right) \,,
\end{equation}
where $\phi_{n_{\mu_\xi}} \!\! \left( \sqrt{\bar{\omega}_\mu}
\,\,\, [{\bm{q}'}^\mu]_\xi \right) $ is a one-dimensional
harmonic-oscillator wave function of frequency
$\bar{\omega}_\mu$ and $n_{\mu_\xi}$ is the oscillator quantum
number, $0 \leq n_{\mu_\xi} < \infty$\,, which counts the number
of quanta in each normal mode. The quantity $\mu$ labels the
manifold of normal modes with the same frequency
$\bar{\omega}_\mu$ while $d_{\mu} = 1$\,, $N-\nolinebreak 1$ or
$N(N-\nolinebreak 3)/2$ for $\mu = {\bf 0}^\pm$\,, ${\bf 1}^\pm$
or ${\bf 2}$ respectively.

\section{Calculating the Normal Mode Coordinates.} \label{sec:NormCalc}
In Paper~I, we extended previous work to the calculation of the
normal coordinates.  We summarize this work below in two-steps:

\newcounter{twostep}
\newcounter{twostepseca}
\newcounter{twostepsecb}
\newcounter{twostepsecc}
\begin{list}{\alph{twostep}).}{\usecounter{twostep}\setlength{\rightmargin}{\leftmargin}}
\item \setcounter{twostepsecb}{\value{twostep}} In Paper I we have
determined two sets of linear combinations of the elements of
coordinate vector $\bar{\bm{r}}'$ which transform under particular
orthogonal $[N]$ and $[N-1, \hspace{1ex} 1]$ irreducible
representations of $S_N$\,. These are the symmetry coordinates for
the $\bar{\bm{r}}'$ sector of the problem\cite{dcw}. Using these
two sets of coordinates we then determined two sets of linear
combinations of the elements of coordinate vector
$\overline{\bm{\gamma}}'$ which transform under exactly the same
orthogonal $[N]$ and $[N-1, \hspace{1ex} 1]$ irreducible
representations of $S_N$ as the coordinate sets in the
$\bar{\bm{r}}'$ sector was determined. Then another set of linear
combinations of the elements of coordinate vector
$\overline{\bm{\gamma}}'$\,, which transforms under a particular
orthogonal $[N-2, \hspace{1ex} 2]$ irreducible representation of
$S_N$\,, was determined. These are then the symmetry coordinates
for the $\overline{\bm{\gamma}}'$ sector of the problem\cite{dcw}.
Furthermore, we chose one of the symmetry coordinates to have the
simplest functional form possible under the requirement that it
transforms irreducibly under $S_N$\,. The succeeding symmetry
coordinate was then chosen to have the next simplest functional
form possible under the requirement that it transforms irreducibly
under $S_N$\,, and so on. In this way the complexity of the
functional form of the symmetry coordinates has been kept to a
minimum and only builds up slowly as more symmetry coordinates of
a given species were considered.
\item  \setcounter{twostepsecc}{\value{twostep}} The $\bm{FG}$
matrix is expressed in the $\bar{\bm{r}}'$\,,
$\overline{\bm{\gamma}}'$ basis. However, if we change the basis
in which the $\bm{FG}$ matrix is expressed to the symmetry
coordinates an enormous simplification occurs. The $N(N+1)/2
\times N(N+1)/2$ eigenvalue equation of Eq.~(\ref{eq:FGit}) is
reduced to one $2 \times 2$ eigenvalue equation for the $[N]$
sector, $N-1$ identical $2 \times 2$ eigenvalue equations for the
$[N-1, \hspace{1ex} 1]$ sector and $N(N-3)/2$ identical $1 \times
1$ eigenvalue equations for the $[N-2, \hspace{1ex} 2]$ sector. In
the case of the $2 \times 2$ eigenvalue equations for the $[N-1,
\hspace{1ex} 1]$ and $[N-2, \hspace{1ex} 2]$ sectors, the $2
\times 2$ structure allows for the mixing in the normal
coordinates of the symmetry coordinates in the $\bar{\bm{r}}'$ and
$\overline{\bm{\gamma}}'$ sectors. The $1 \times 1$ structure of
the eigenvalue equations in the $[N-2, \hspace{1ex} 2]$ sector
reflects the fact that there are no $[N-2, \hspace{1ex} 2]$
symmetry coordinates in the $\bar{\bm{r}}'$ sector for the $[N-2,
\hspace{1ex} 2]$ symmetry coordinates in the
$\overline{\bm{\gamma}}'$ sector to couple with. The $[N-2,
\hspace{1ex} 2]$ normal modes are entirely angular.
\end{list}

\subsection{Transformation to Symmetry Coordinates}
Let the symmetry coordinate vector,
${\bm{S}}$\,, be
\begin{equation}\label{eq:FSCV}
{\bm{S}} = \left( \begin{array}{l} {\bm{S}}_{\bar{\bm{r}}'}^{[N]} \\
{\bm{S}}_{\overline{\bm{\gamma}}'}^{[N]}  \\ \hline {\bm{S}}_{\bar{\bm{r}}'}^{[N-1, \hspace{1ex} 1]} \\
{\bm{S}}_{\overline{\bm{\gamma}}'}^{[N-1, \hspace{1ex} 1]} \\
\hline
{\bm{S}}_{\overline{\bm{\gamma}}'}^{[N-2, \hspace{1ex} 2]} \end{array} \right) = \left( \begin{array}{l} {\bm{S}}^{[N]} \\
{\bm{S}}^{[N-1, \hspace{1ex} 1]} \\
{\bm{S}}^{[N-2, \hspace{1ex} 2]} \end{array} \right).
\end{equation}
%
%
%
In ${\bm{S}}$\,, symmetry coordinates of the same species are
grouped together. From Paper I we have
\renewcommand{\arraystretch}{3}
\begin{equation} \label{eq:SNeqsqrtN1}
\begin{array}{rcl}
{\displaystyle {\bm{S}}_{\bar{\bm{r}}'}^{[N]} } & = &
{\displaystyle \frac{1}{\sqrt{N}} \,
\sum_{i=1}^N \overline{r}'_i \,, } \\
{\displaystyle {\bm{S}}_{\overline{\bm{\gamma}}'}^{[N]} } & = &
{\displaystyle \sqrt{\frac{2}{N(N-1)}} \,\,\, \sum_{j=2}^N \sum_{i
< j} \overline{\gamma}'_{ij} \,, } \mbox{\hspace{1em}and}
\end{array}
\end{equation}
\renewcommand{\arraystretch}{1}
%
%
%
%
\vspace{-0.8em}
\begin{eqnarray} \label{eq:SNm1}
[{\bm{S}}_{\bar{\bm{r}}'}^{[N-1, \hspace{1ex} 1]}]_i & = &
\frac{1}{\sqrt{i(i+1)}} \left(
\sum_{k=1}^i \bar{r}'_k - i \bar{r}'_{i+1} \right)\,, \\
&& \;\;\; \mbox{where} \;\;\; 1 \leq i \leq N-1 \,. \nonumber
\end{eqnarray}
We see from Eq.~(\ref{eq:SNm1}) the first symmetry coordinate is
proportional to  $\bar{r}'_1 - \bar{r}'_2$\,. This involves
only the first two particles in the simplest motion possible under
the requirement that the symmetry coordinate transforms
irreducibly under the $[N-1, \hspace{1ex} 1]$ representation of
$S_N$\,. We again advert that adding another particle to the
system does not cause widespread disruption to the symmetry
coordinates. The symmetry coordinates, and the motions they
describe, remain the same except for an additional symmetry
coordinate which describes a motion involving all of the particles.

Again according to item~\alph{twostepsecb}).\ above,
${\bm{S}}_{\overline{\bm{\gamma}}'}^{[N-1, \hspace{1ex} 1]}$
should transform under exactly the same non-orthogonal irreducible
$[N-1, \hspace{1ex} 1]$ representation of $S_N$ as
${\bm{S}}_{\bar{\bm{r}}'}^{[N-1, \hspace{1ex} 1]}$\,. From Paper I
\begin{widetext}
\begin{equation} \label{eq:SgNm1i}
\hspace*{-0.5em} \renewcommand{\arraystretch}{2}
\begin{array}{rcl}
[{\bm{S}}_{\overline{\bm{\gamma}}'}^{[N-1, \hspace{1ex} 1]}]_i & =
& {\displaystyle \frac{1}{\sqrt{i(i+1)(N-2)}} \, \left(
\sum_{k=1}^i \sum_{l=1}^N
\overline{\gamma}'_{kl} - i \sum_{l=1}^N \overline{\gamma}'_{i+1,\,l} \right) } \\
& = & {\displaystyle \frac{1}{\sqrt{i(i+1)(N-2)}} \, \left( \left[
\sum_{l = 2}^i \, \sum_{k=1}^{l-1} \overline{\gamma}'_{kl} +
\sum_{k = 1}^i \, \sum_{l=k+1}^{N} \hspace{-1ex}
\overline{\gamma}'_{kl} \right] - i \left[ \sum_{k=1}^i
\overline{\gamma}'_{k,\,i+1} + \sum_{l=i+2}^N \hspace{-0.5ex}
\overline{\gamma}'_{i+1,\,l} \right] \right) \,. }
\end{array} \renewcommand{\arraystretch}{1}
\end{equation}
\end{widetext}
where $  1 \leq i \leq N-1 \,$.
Now there is only one sector, the $\overline{\bm{\gamma}}'$
sector, belonging to the $[N-2, \hspace{1ex} 2]$ species. From
Paper I
\begin{widetext}
\begin{equation} \label{eq:SNm2gip1j}
[{\bm{S}}_{\overline{\bm{\gamma}}'}^{[N-2, \hspace{1ex} 2]}]_{ij}
 = \frac{1}{\sqrt{i(i+1)(j-3)(j-2)}} \, \left(
\vphantom{\sum_{k=1}^{[j'-1, i]_{min}} \hspace{-2ex}
\bar{\gamma}'_{kj'}} \right. 
\begin{array}[t]{@{}c@{}}
\displaystyle{ \sum_{j'=2}^{j-1} \sum_{k=1}^{[j'-1, i]_{min}}
\hspace{-2ex} \bar{\gamma}'_{kj'} + \sum_{k=1}^{i-1}
\sum_{j'=k+1}^i
\bar{\gamma}'_{kj'} - (j-3) \sum_{k=1}^i \bar{\gamma}'_{kj} - } \\
\left. \displaystyle{ - i \sum_{k=1}^{i} \bar{\gamma}'_{k,(i+1)} -
i \sum_{j'=i+2}^{j-1} \bar{\gamma}'_{(i+1),j'} + i (j-3)
\bar{\gamma}'_{(i+1),j} } \right) \,, \end{array}
\end{equation}
\end{widetext}
where $1 \leq i \leq j-2$ and $4 \leq j \leq N$\,.

This fulfills item~\alph{twostepsecb}).\ above: the determination
of the symmetry coordinates.

\subsection{Transformation to Normal-Mode Coordinates.} \label{sec:FreqNorModN}
In Paper~I we apply the $\bm{FG}$ method using these symmetry
coordinates to determine the frequencies and normal modes of the
system:
\begin{equation} \label{eq:lambda12pm}
\lambda^\pm_\alpha = \frac{a_\alpha \pm \sqrt{b_\alpha^2 +
4\,c_\alpha }}{2}
\end{equation}
for the $\alpha=[N]$ and $[N-1, \hspace{1ex} 1]$ sectors, where
\begin{eqnarray}
a_\alpha & = &
\protect[\bm{\sigma_\alpha^{FG}}\protect]_{\bar{\bm{r}}',\,\bar{\bm{r}}'}
+
\protect[\bm{\sigma_\alpha^{FG}}\protect]_{\overline{\bm{\gamma}}',\,\overline{\bm{\gamma}}'}
\nonumber \\
b_\alpha & = &
\protect[\bm{\sigma_\alpha^{FG}}\protect]_{\bar{\bm{r}}',\,\bar{\bm{r}}'}
-
\protect[\bm{\sigma_\alpha^{FG}}\protect]_{\overline{\bm{\gamma}}',\,\overline{\bm{\gamma}}'}
\\ c_\alpha & = & \protect[\bm{\sigma_\alpha^{FG}}\protect]_{\bar{\bm{r}}',\,
\overline{\bm{\gamma}}'} \, \times
\protect[\bm{\sigma_\alpha^{FG}}\protect]_{\overline{\bm{\gamma}}',\,\bar{\bm{r}}'}
\nonumber \,,
\end{eqnarray}
while $\lambda_{[N-2, \hspace{1ex} 2]} = \bm{\sigma_{[N-2,
\hspace{1ex} 2]}^{FG}}$\,. The $\bm{\sigma_\alpha^{FG}}$ are
related to the $\tilde{a}$\,, $\tilde{b}$\,, $\tilde{c}$\,,
$\tilde{d}$\,, $\tilde{e}$\,, $\tilde{f}$\,, $\tilde{g}$\,,
$\tilde{h}$\,, and $\tilde{\iota}$ of the $\bm{FG}$ matrix of
Eq.~(\ref{eq:GFsub}). Note that the $\bm{\sigma_\alpha^{FG}}$ for
the $\alpha=[N]$ and $[N-1, \hspace{1ex} 1]$ sectors are $2 \times
2$ matrices while $\bm{\sigma_{[N-2, \hspace{1ex} 2]}^{FG}}$ is a
one-component quantity.

The normal coordinates are given by
\begin{equation} \label{eq:qnpfullexp}
{\bm{q}'}_\pm^\alpha = c_\pm^{\alpha} \left(
\cos{\theta^\alpha_\pm} \, [{\bm{S}}_{\bar{\bm{r}}'}^{\alpha}]_\xi
\, + \, \sin{\theta^\alpha_\pm} \,
[{\bm{S}}_{\overline{\bm{\gamma}}'}^{\alpha}]_\xi \right)
\end{equation}
for the $\alpha=[N]$ and $[N-1, \hspace{1ex} 1]$ sectors,
and
\begin{equation} \label{eq:qnm2fullexp}
{\bm{q}'}^{[N-2, \hspace{1ex} 2]} = c^{[N-2, \hspace{1ex} 2]}
{\bm{S}}_{\overline{\bm{\gamma}}'}^{[N-2, \hspace{1ex} 2]} \,.
\end{equation}
The
$\bar{\bm{r}}'$-$\overline{\bm{\gamma}}'$ mixing angle,
$\theta^\alpha_\pm$\,, is given by
\begin{equation} \label{eq:tanthetaalphapm}
\tan{\theta^\alpha_\pm} = \frac{(\lambda^\pm_\alpha -
\protect[\bm{\sigma_\alpha^{FG}}\protect]_{\bar{\bm{r}}',\,\bar{\bm{r}}'})}
{\protect[\bm{\sigma_\alpha^{FG}}\protect]_{\bar{\bm{r}}',\,
\overline{\bm{\gamma}}'}} =
\frac{\protect[\bm{\sigma_\alpha^{FG}}\protect]_{\overline{\bm{\gamma}}',\,\bar{\bm{r}}'}}{(\lambda^\pm_\alpha
-
\protect[\bm{\sigma_\alpha^{FG}}\protect]_{\overline{\bm{\gamma}}',\,\overline{\bm{\gamma}}'})}\,,
\end{equation}
while the normalization constants $c$ are given by
\begin{equation} \label{eq:calphapm}
\begin{array}{l}
c_\pm^\alpha   = {\displaystyle  \frac{1}{\sqrt{\left(
\begin{array}{c} \cos{\theta^\alpha_\pm} \\
\sin{\theta^\alpha_\pm} \end{array} \right)^T
\bm{\sigma_{\alpha}^{G}} \left( \begin{array}{c}
\cos{\theta^\alpha_\pm} \\ \sin{\theta^\alpha_\pm} \end{array}
\right)}} \mbox{\hspace{1em}and\hspace{1em}} } \\ \\
{\displaystyle c^{[N-2, \hspace{1ex} 2]} =
\frac{1}{\sqrt{\bm{\sigma_{[N-2, \hspace{1ex} 2]}^G}}} } \,\,.
\end{array}
\end{equation}
The $\bm{\sigma_{\alpha}^{G}}$ are related to the $\tilde{a}'$\,,
$\tilde{g}'$\,, and $\tilde{h}'$ of the $\bm{G}$ matrix of
Eq.~(\ref{eq:GFsub}).

This fulfills item~\alph{twostepsecc}).\ above; the determination
of the frequencies and normal coordinates.
Equations~(\ref{eq:qnpfullexp}) and (\ref{eq:qnm2fullexp})
represent the final results of Paper~I, the normal coordinates.

\section{Deriving the Density Profile of a BEC from the Wave
Function} \label{sec:densityprofile} As a first application of this
general theory from Paper~I, we calculate the density profile for a
confined quantum system. The harmonic-order DPT wave function,
$_g\!\Phi_0({\bar{\bm{y}}}^{\prime})$\,, for the ground state is
given by Eq.~(\ref{eq:wavefunct}) with all of the $n_{\mu_\xi}$
set equal to zero, i.e.\
\begin{equation} \label{eq:groundWF}
_g\!\Phi_0({\bar{\bm{y}}}^{\prime}) = \prod_{ \mu = \{
\bm{0}^\pm,\hspace{0.5ex} \bm{1}^\pm,\hspace{0.5ex} \bm{2} \} }
 \,\, \prod_{\xi=1}^{d_{\mu}} \,\, \phi_{0}
\!\! \left( \sqrt{\bar{\omega}_\mu} \,\,\, [{\bm{q}'}^\mu]_\xi
\right) \,,
\end{equation}
where
\begin{equation} \label{eqphi0}
\phi_{0} \!\! \left( \sqrt{\bar{\omega}_\mu} \,\,\,
[{\bm{q}'}^\mu]_\xi \right) = \left( \frac{\bar{\omega}_\mu}{\pi}
\right)^{\frac{1}{4}} \,\, \exp{\left(-\frac{1}{2}
\bar{\omega}_\mu [{\bm{q}'}^\mu]^2_\xi \right) } \,.
\end{equation}
Now the ground-state wave function of a BEC is completely
symmetric under interchange of any of the particles and so we
require $_g\!\Phi_0({\bar{\bm{y}}}^{\prime})$ of
Eq.~(\ref{eq:groundWF}) to be completely symmetric under
interchange of any of the particles. This follows from the fact
that the five \, $\prod_{\xi=1}^{d_{\mu}} \,\, \phi_{0} \!\!
\left( \sqrt{\bar{\omega}_\mu} \,\,\, [{\bm{q}'}^\mu]_\xi
\right)$\,, where $\mu = \{ \bm{0}^\pm,\hspace{0.5ex}
\bm{1}^\pm,\hspace{0.5ex} \bm{2} \}$\,, are each completely
symmetric under interchange of any of the particles. This may be
seen from the fact that
\begin{eqnarray}
\lefteqn{\hspace{-2ex} \prod_{\xi=1}^{d_{\mu}} \,\, \phi_{0} \!\!
\left( \sqrt{\bar{\omega}_\mu} \,\,\, [{\bm{q}'}^\mu]_\xi \right)
} \nonumber \\ & = & \left( \sqrt{\frac{\bar{\omega}_\mu}{\pi}}
\right)^{d_{\mu}} \exp{\left(-\frac{1}{2} \bar{\omega}_\mu \left\{
\sum_{\xi=1}^{d_{\mu}} \, [{\bm{q}'}^\mu]^2_\xi \right\} \right)}
\,.  \label{eq:phi0funct}
\end{eqnarray}
Now according to items~\alph{twostepsecb}).\ and
\alph{twostepsecc}).\ of Sec.~\ref{sec:NormCalc}, particle
interchange is effected by orthogonal transformations of the
$[{\bm{q}'}^\mu]_\xi$\,, where $1 \leq \xi \leq d_{\mu}$\,, and
any orthogonal transformation leaves $\sum_{\xi=1}^{d_{\mu}} \,
[{\bm{q}'}^\mu]^2_\xi$ invariant. Together with
Eq.~(\ref{eq:phi0funct}), this fact means that the
$\prod_{\xi=1}^{d_{\mu}} \,\, \phi_{0} \!\! \left(
\sqrt{\bar{\omega}_\mu} \,\,\, [{\bm{q}'}^\mu]_\xi \right)$ are
completely symmetric under interchange of any of the particles as
advertised.

Defining $S(D)$ to be the total $D$-dimensional solid
angle\cite{hypharm},
\begin{equation}
S(D)=\frac{2 \,\, \pi^{\frac{D}{2}}}{\Gamma(\frac{D}{2})} \,,
\end{equation}
where we note that $S(1)=2$\,, $S(2)=2\pi$\,, $S(3)=4\pi$\,,
$S(4)=2\pi^2$\,, \ldots\,, the harmonic-order Jacobian-weighted
ground-state density profile, $N_0(r)$\,, is
\begin{equation}
\renewcommand{\arraystretch}{1.5}
\begin{array}[b]{@{}l@{}}
{\displaystyle S(D) \, N_0(r) = S(D) \, r^{(D-1)} \rho_0(r) } \\
{\displaystyle =  \sum_{i=1}^N \int_{-\infty}^\infty \cdots
\int_{-\infty}^\infty \hspace{-1.3ex} \delta_f(r-r_i) \,
[_g\!\Phi_0({\mathbf{\bar{y}'}})]^2 \hspace{-2ex} \prod_{\mu= {\bf
0}^\pm, {\bf 1}^\pm, {\bf 2}} \, \prod_{\xi=1}^{d_{\mu}}
d[{\mathbf{q}'}^\mu]_\xi } \,,
\end{array}
\renewcommand{\arraystretch}{1}
\end{equation}
where $\rho_0(r)$ is the unweighted harmonic-order ground-state
density profile and $\delta_f(r-r_i)$ is the Dirac delta function
and is differentiated from the inverse dimension, $\delta$\,, by
the subscript $f$\,. Since $_g\!\Phi_0({\bar{\bm{y}}}^{\prime})$
is invariant under particle interchange then
\begin{equation}
\renewcommand{\arraystretch}{1.5}
\begin{array}[b]{@{}l@{}}
{\displaystyle S(D) \, N_0(r) } \\
 {\displaystyle =  N \int_{-\infty}^\infty \hspace{-0.5ex} \cdots \int_{-\infty}^\infty
\hspace{-1.3ex} \delta_f(r-r_N) \,
[_g\!\Phi_0({\mathbf{\bar{y}'}})]^2 \hspace{-2ex} \prod_{\mu= {\bf
0}^\pm, {\bf 1}^\pm, {\bf 2}} \, \prod_{\xi=1}^{d_{\mu}}
d[{\mathbf{q}'}^\mu]_\xi \,.}
\end{array}
\renewcommand{\arraystretch}{1}
\end{equation}
Upon using Eq.~(\ref{eq:groundWF}) and the fact that $r_{N}$ only
appears in ${\bm{q}'}^{{\bf 0}^+}$\,, ${\bm{q}'}^{{\bf 0}^-}$\,,
$[{\bm{q}'}^{{\bf 1}^+}]_{d_{{\bf 1}^+}}$ and $[{\bm{q}'}^{{\bf
1}^-}]_{d_{{\bf 1}^-}}$ (see Eqs.~(\ref{eq:kappascale}),
(\ref{eq:taylor1}), (\ref{eq:SNeqsqrtN1})-(\ref{eq:SNm2gip1j}),
(\ref{eq:qnpfullexp}), (\ref{eq:qnm2fullexp})) and
Eq.~(\ref{eqphi0})) we obtain
\begin{equation}
\renewcommand{\arraystretch}{1.5}
\begin{array}[b]{@{}l@{}}
{\displaystyle S(D) \, N_0(r) } \\
 = \begin{array}[t]{@{}l@{}} {\displaystyle N \int_{-\infty}^\infty
\int_{-\infty}^\infty \int_{-\infty}^\infty
\int_{-\infty}^\infty \delta_f(r-r_N) } \\
{\displaystyle \hspace{13ex} \times \hspace{-1ex} \prod_{\mu= {\bf
0}^\pm, {\bf 1}^\pm} \, [\, \phi_0 \! \left(
\sqrt{\bar{\omega}_\mu} \, [{\mathbf{q}'}^\mu]_{d_{\mu}}
\right)]^2 \, d[{\mathbf{q}'}^\mu]_{d_{\mu}} } \end{array}
\\ \\ =  \begin{array}[t]{@{}l@{}}
{\displaystyle \frac{N \sqrt{ \bar{\omega}_{{\bf 0}^+} \,
\bar{\omega}_{{\bf 0}^-} \, \bar{\omega}_{{\bf 1}^+} \,
\bar{\omega}_{{\bf 1}^-} } }{\pi^2} \! \int_{-\infty}^\infty
\int_{-\infty}^\infty \int_{-\infty}^\infty
\int_{-\infty}^\infty \!\!\!\! \delta_f(r-r_N) } \\
{\displaystyle \hspace{4ex} \times \exp{\left(- \hspace{-2ex}
\sum_{\nu= {\bf 0}^\pm, {\bf 1}^\pm} \hspace{-1ex}
\bar{\omega}_\nu [{\bm{q}'}^\nu]^2_{d_{\nu}} \right) } \!
\prod_{\zeta= {\bf 0}^\pm, {\bf 1}^\pm} \hspace{-2ex}
d[{\bm{q}'}^\zeta]_{d_{\zeta}} } \,. \end{array}
\end{array}
\renewcommand{\arraystretch}{1}
\end{equation}
The delta function, $\delta_f(r-r_N)$\,, is a function of $r_N$
while the integral is over the normal coordinates ${\bm{q}'}^{{\bf
0}^+}$\,, ${\bm{q}'}^{{\bf 0}^-}$\,, $[{\bm{q}'}^{{\bf
1}^+}]_{N-1}$\,, and $[{\bm{q}'}^{{\bf 1}^-}]_{N-1}$\,. Thus we need
a change of variables to perform the integral. We change the
variables of the integral to $\bar{r}'_{N}$\,, $\bar{r}'_{S}$\,,
${\bm{S}}_{\overline{\bm{\gamma}}'}^{[N]}$\,, and
$[{\bm{S}}_{\overline{\bm{\gamma}}'}^{[N-1, \hspace{1ex}
1]}]_{(N-1)}$\,, where
\begin{equation} \label{eq:rs}
\bar{r}'_{S} = \sum_{i=1}^{N-1} \bar{r}'_{i} \,,
\end{equation}
${\bm{S}}_{\overline{\bm{\gamma}}'}^{[N]}$ is given by
Eq.~(\ref{eq:SNeqsqrtN1}) and
$[{\bm{S}}_{\overline{\bm{\gamma}}'}^{[N-1, \hspace{1ex}
1]}]_{(N-1)}$ is given by Eq.~(\ref{eq:SgNm1i}). Thus from
Eqs.~(\ref{eq:SNm1}), (\ref{eq:qnpfullexp}),
(\ref{eq:qnm2fullexp}), and (\ref{eq:rs}) we obtain
\begin{equation} \left( \begin{array}{c} {\bm{q}'}^{{\bf 0}^+}
\\ {\bm{q}'}^{{\bf 0}^-} \\ \protect[{\bm{q}'}^{{\bf
1}^+}\protect]_{N-1} \\ \protect[{\bm{q}'}^{{\bf
1}^-}\protect]_{N-1} \end{array} \right) = \, \bm{T} \,\,
{\bm{a}'} \,,
\end{equation}
where 
\renewcommand{\arraystretch}{1.5}
\begin{eqnarray} \label{eq:ap}
{\bm{a}'} & = & \left( \begin{array}{c} \bar{r}'_{N} \\
\bar{r}'_{S}
\\ {\bm{S}}_{\overline{\bm{\gamma}}'}^{[N]} \\ \protect[{\bm{S}}_{\overline{\bm{\gamma}}'}^{[N-1, \hspace{1ex}
1]}\protect]_{(N-1)} \end{array} \right) \,, \nonumber \\ && \\
\bm{T} & = & \left(
\begin{array}{cccc}
T_{11} & T_{11} & T_{13}
& 0 \\
T_{21} & T_{21} & T_{23}
& 0 \\
-(N-1)T_{31} & T_{31} & 0 & T_{34} \\
-(N-1)T_{41} & T_{41} & 0 & T_{44}
\end{array}
\right) \,,
\nonumber
\end{eqnarray}
\renewcommand{\arraystretch}{1}
$\protect\mbox{and}$
\protect\begin{widetext} \protect\renewcommand{\jot}{0.5em}
\protect\begin{eqnarray} T_{11} & = & \frac{c^{[N]}_+
\cos{\theta^{[N]}_+} }{\sqrt{N}} \,, \hspace{1em} T_{13} =
c^{[N]}_+ \sin{\theta^{[N]}_+} \,, \hspace{1em} T_{21} =
\frac{c^{[N]}_- \cos{\theta^{[N]}_-} }{\sqrt{N}} \,,
\hspace{1em} T_{23} = c^{[N]}_- \sin{\theta^{[N]}_-} \,, \nonumber \\
T_{31} & = & \frac{c^{[N-1, \hspace{1ex} 1]}_+ \cos{\theta^{[N-1,
\hspace{1ex} 1]}_+} }{\sqrt{N(N-1)}}  \,, \hspace{1em} T_{34} =
c^{[N-1, \hspace{1ex} 1]}_+ \sin{\theta^{[N-1,
\hspace{1ex} 1]}_+} \,, \\
T_{41} & = & \frac{c^{[N-1, \hspace{1ex} 1]}_- \cos{\theta^{[N-1,
\hspace{1ex} 1]}_-} }{\sqrt{N(N-1)}}  \,, \hspace{1em} T_{44} =
c^{[N-1, \hspace{1ex} 1]}_- \sin{\theta^{[N-1, \hspace{1ex} 1]}_-}
\,. \nonumber \protect\end{eqnarray}
\protect\renewcommand{\jot}{0em} \protect\end{widetext}
The Jacobian, $J_T$\,, of the transformation is thus
\begin{eqnarray}
\lefteqn{ \hspace{-2ex} J_T = \det{T} = - \frac{c^{[N]}_+ \,
c^{[N]}_- \, c^{[N-1, \hspace{1ex} 1]}_+ \, c^{[N-1, \hspace{1ex}
1]}_-}{\sqrt{N-1}} } \nonumber \\ \hspace{-2ex} && \times \,
\sin{(\theta^{[N]}_+ - \theta^{[N]}_-)} \, \sin{(\theta^{[N-1,
\hspace{1ex} 1]}_+ - \theta^{[N-1, \hspace{1ex} 1]}_-)} \,.
\end{eqnarray}
Defining
\begin{equation}
{\bm{\overline{\Omega}}} = \left( \begin{array}{cccc}
\bar{\omega}_{{\bf 0}^+} & 0 & 0 & 0 \\ 0 & \bar{\omega}_{{\bf
0}^-} & 0 & 0 \\ 0 & 0 & \bar{\omega}_{{\bf 1}^+} & 0 \\ 0 & 0 & 0
& \bar{\omega}_{{\bf 1}^-}
\end{array} \right)
\end{equation}
then
\begin{equation} \label{eq:Nragain}
\renewcommand{\arraystretch}{2}
\begin{array}[b]{@{}l@{}}
{\displaystyle S(D) \, N_0(r) } \\
= \begin{array}[t]{@{}l@{}} {\displaystyle \frac{N \, J_T \sqrt{
\bar{\omega}_{{\bf 0}^+} \, \bar{\omega}_{{\bf 0}^-} \,
\bar{\omega}_{{\bf 1}^+} \, \bar{\omega}_{{\bf 1}^-} } }{\pi^2} \!
\int_{-\infty}^\infty \int_{-\infty}^\infty \int_{-\infty}^\infty
\int_{-\infty}^\infty \!\!\!\! \delta_f(r-r_N) } \\
{\displaystyle \hspace{13ex} \times \hspace{0ex} \exp{\left(-
{\bm{a}'}^T \bm{T}^T {\bm{\overline{\Omega}}} \, \bm{T} {\bm{a}'}
\right) } \, d\bar{r}'_{N} \, d^3{\bm{b}'} } \,,
\end{array} \end{array}
\renewcommand{\arraystretch}{1}
\end{equation}
where $d^3{\bm{b}'} = d\bar{r}'_{S} \,
d{\bm{S}}_{\overline{\bm{\gamma}}'}^{[N]} \,
d[{\bm{S}}_{\overline{\bm{\gamma}}'}^{[N-1, \hspace{1ex}
1]}]_{(N-1)}$\,. Writing
\begin{equation} \label{eq:TK}
\bm{T}^T  {\bm{\overline{\Omega}}} \, \bm{T} = \left(
\begin{array}{cc} K_0 & \bm{K}^T \\
\bm{K} & \bm{\mathcal{K}}
\end{array} \right) \,,
\end{equation}
where
\begin{equation}
K_0 = \bar{\omega}_{{\bf 0}^+} T_{11}^2 +
    \bar{\omega}_{{\bf 0}^-} T_{21}^2 + (N-1)^2 (\bar{\omega}_{{\bf 1}^+} T_{31}^2 +
          \bar{\omega}_{{\bf 1}^-} T_{41}^2) \,,
\end{equation}
%
%
\begin{widetext}
\renewcommand{\jot}{0.5em}
\begin{eqnarray}
\bm{K} & = & \left( \begin{array}{c} \bar{\omega}_{{\bf 0}^+}
T_{11}^2 +
    \bar{\omega}_{{\bf 0}^-} T_{21}^2 - (N-1) (\bar{\omega}_{{\bf 1}^+} T_{31}^2 +
          \bar{\omega}_{{\bf 1}^-} T_{41}^2)  \\  \bar{\omega}_{{\bf 0}^+} T_{11} T_{13}
          + \bar{\omega}_{{\bf 0}^-} T_{21} T_{23}  \\
          -(N-1) (\bar{\omega}_{{\bf 1}^+} T_{31} T_{34} +
          \bar{\omega}_{{\bf 1}^-} T_{41} T_{44})  \end{array}
\right) \,, \\
\bm{\mathcal{K}}_{11} & = & \bar{\omega}_{{\bf 0}^+} T_{11}^2 +
\bar{\omega}_{{\bf 0}^-} T_{21}^2 + \bar{\omega}_{{\bf 1}^+}
T_{31}^2 + \bar{\omega}_{{\bf 1}^-} T_{41}^2 \,, \hspace{1em}
\bm{\mathcal{K}}_{12} = \bm{\mathcal{K}}_{21} = \bar{\omega}_{{\bf
0}^+} T_{11} T_{13} + \bar{\omega}_{{\bf 0}^-}
T_{21} T_{23} \,, \nonumber \\
\bm{\mathcal{K}}_{13} = \bm{\mathcal{K}}_{31} & = &
\bar{\omega}_{{\bf 1}^+} T_{31} T_{34} + \bar{\omega}_{{\bf 1}^-}
T_{41} T_{44} \,, \hspace{1em} \bm{\mathcal{K}}_{22} =
\bar{\omega}_{{\bf 0}^+} T_{13}^2 +
\bar{\omega}_{{\bf 0}^-} T_{23}^2 \,, \\
\bm{\mathcal{K}}_{23} = \bm{\mathcal{K}}_{32} & = & 0 \,,
\hspace{1em} \bm{\mathcal{K}}_{33} = \bar{\omega}_{{\bf 1}^+}
T_{34}^2 + \bar{\omega}_{{\bf 1}^-} T_{44}^2 \,. \nonumber
\end{eqnarray}
\renewcommand{\jot}{0em}
\end{widetext}
Using Eqs.~(\ref{eq:ap}) and (\ref{eq:TK}) in
Eq.~(\ref{eq:Nragain}), we obtain
\begin{equation} 
\renewcommand{\arraystretch}{2}
\begin{array}[b]{@{}l@{}}
{\displaystyle S(D) \, N_0(r) } \\
= \begin{array}[t]{@{}l@{}} {\displaystyle \frac{N \, J_T \sqrt{
\bar{\omega}_{{\bf 0}^+} \, \bar{\omega}_{{\bf 0}^-} \,
\bar{\omega}_{{\bf 1}^+} \, \bar{\omega}_{{\bf 1}^-} } }{\pi^2} \!
\int_{-\infty}^\infty \int_{-\infty}^\infty \int_{-\infty}^\infty
\int_{-\infty}^\infty \!\!\!\! \delta_f(r-r_N) } \\
{\displaystyle \hspace{2ex} \times \hspace{0ex} \exp{\left(- K_0
\bar{r}_{N}^{\prime 2} - 2 \bar{r}'_{N} \bm{K}^T \bm{b}' -
\bm{b}'^T \bm{\mathcal{K}} \bm{b}' \right) } \, d\bar{r}'_{N} \,
d^3{\bm{b}'} \,, }
\end{array} \end{array}
\renewcommand{\arraystretch}{1}
\end{equation}
where
\begin{equation}
\bm{b}' = \left( \begin{array}{c} \bar{r}'_{S}
\\ {\bm{S}}_{\overline{\bm{\gamma}}'}^{[N]} \\ \protect[{\bm{S}}_{\overline{\bm{\gamma}}'}^{[N-1, \hspace{1ex}
1]}\protect]_{(N-1)} \end{array} \right) \,.
\end{equation}
Since $\delta_f(r-r_N) = \sqrt{\delta} \,\, \kappa(D) \,\,
\delta_f\left(\delta^{-\frac{1}{2}}\left(\frac{r}{\kappa(D)}
 - \bar{r}_{\infty}\right) - \bar{r}'_{N}\right)$\,, then
\begin{eqnarray}
\lefteqn{S(D) \, N_0(r) = \frac{N \, J_T \sqrt{\delta} \,\,
\kappa(D) \,\, \sqrt{ \bar{\omega}_{{\bf 0}^+} \,
\bar{\omega}_{{\bf 0}^-} \, \bar{\omega}_{{\bf 1}^+} \,
\bar{\omega}_{{\bf 1}^-} } }{\pi^2} } \nonumber \\ 
&& \times \int_{-\infty}^\infty \int_{-\infty}^\infty
\int_{-\infty}^\infty \int_{-\infty}^\infty
\exp{
\left(- \delta^{-1} K_0 \left(\frac{r}{\kappa(D)} -
\bar{r}_{\infty}\right)^2 \right. } \nonumber \\
&& \hspace{4ex} \left. \vphantom{- \delta^{-1} K_0
\left(\frac{r}{\kappa(D)} - \bar{r}_{\infty}\right)^2} - 2
\delta^{-\frac{1}{2}}\left(\frac{r}{\kappa(D)}
 - \bar{r}_{\infty}\right) \bm{K}^T \bm{b}' - \bm{b}'^T \bm{\mathcal{K}} \bm{b}'
\right)
\, d^3{\bm{b}'} \,. \nonumber \\  \label{eq:Nrll}
\end{eqnarray}
Using
\begin{eqnarray} \label{eq:multidimgaussintegral}
\lefteqn{ \hspace{-10ex} \int_{-\infty}^\infty \cdots
\int_{-\infty}^\infty \exp{\left( - \bm{b}'^T \bm{A} \bm{b}' -  2
\bm{B}^T \bm{b}' \right) } \, d^n{\bm{b}'} } \nonumber \\ &
\hspace{4ex} = & \frac{\pi^{\frac{n}{2}}}{\sqrt{\det{\bm{A}}}}
\exp{\left( \bm{B}^T \bm{A}^{-1} \bm{B} \right) } \,,
\end{eqnarray}
 Eq.~(\ref{eq:Nrll}) yields
\begin{eqnarray} 
\lefteqn{ \hspace{-3ex} S(D) \, N_0(r) = \frac{N \sqrt{\delta}
\,\, \kappa(D) \,\,}{\sqrt{\pi}} \, \sqrt{ \frac{
\bar{\omega}_{{\bf 0}^+} \, \bar{\omega}_{{\bf 0}^-} \,
\bar{\omega}_{{\bf 1}^+} \, \bar{\omega}_{{\bf 1}^-} \, J_T^2 }{
\det{\bm{\mathcal{K}}}} } } \nonumber \\ && \hspace{-4ex} \times
\exp{\left(- \delta^{-1} \left(\frac{r}{\kappa(D)} -
\bar{r}_{\infty}\right)^2
 (K_0 - \bm{K}^T \bm{\mathcal{K}}^{-1} \bm{K} ) \right) } \,.
 \nonumber \\
\end{eqnarray}
Since
\begin{equation}
\frac{ \bar{\omega}_{{\bf 0}^+} \, \bar{\omega}_{{\bf 0}^-} \,
\bar{\omega}_{{\bf 1}^+} \, \bar{\omega}_{{\bf 1}^-} \, J_T^2 }{
\det{\bm{\mathcal{K}}}} = (K_0 - \bm{K}^T \bm{\mathcal{K}}^{-1}
\bm{K} ) \,,
\end{equation}
then
\begin{eqnarray} \label{eq:Nrresult}
\lefteqn{ \hspace{-3ex} S(D) \, N_0(r) = N \, \sqrt{ \frac{\delta
\, \kappa^2(D) \,\, (K_0 - \bm{K}^T \bm{\mathcal{K}}^{-1} \bm{K} )
}{\pi} } } \nonumber \\ && \hspace{-4ex} \times \exp{\left(-
\delta^{-1} \left(\frac{r}{\kappa(D)} - \bar{r}_{\infty}\right)^2
 (K_0 - \bm{K}^T \bm{\mathcal{K}}^{-1} \bm{K} ) \right) } \,.
 \nonumber \\
\end{eqnarray}

The harmonic-order DPT Jacobian-weighted density profile,
$N_0(r)$\,, is correctly normalized to $N$\,, since upon using
Eq.~(\ref{eq:multidimgaussintegral}) (with $n=1$\,, $\bm{A} = (K_0
- \bm{K}^T \bm{\mathcal{K}}^{-1} \bm{K} ) / (\delta \kappa^2(D))$
and $\bm{B}= - \bar{r}_{\infty} \, (K_0 - \bm{K}^T
\bm{\mathcal{K}}^{-1} \bm{K} ) \ (\delta \kappa(D))$) in
Eq.~(\ref{eq:Nrresult}) we find
\begin{equation}
\int_\Omega \int_{-\infty}^\infty N_0(r) \, dr \, d\Omega = N \,.
\end{equation}
Notice that the harmonic-order DPT density profile is a normalized
gaussian centered around $r=\kappa(D) \, \bar{r}_{\infty}$\,, the
 $D\rightarrow\infty$
configuration radius (see Eqs.~(\ref{eq:kappascale}) and
(\ref{eq:Lewis})).

\section{Optimizing the Harmonic-Order Result} \label{sec:optimize}
Let's look in more detail at the harmonic-order wave function. In
the notation of Eqs.~(\ref{eq:ytranspose}) and
(\ref{eq:ytransposeP}), Eq.~(\ref{eq:taylor1}) can be written as
\begin{equation}
{\bar{\bm{y}}} = {\bar{\bm{y}}}_\infty + \delta^{1/2}
{\bar{\bm{y}}}^{\prime} \,, \label{eq:ytoy}
\end{equation}
where
\begin{equation}
{\bar{\bm{y}}}_\infty =
\begin{array}[t]{r|l} {\bar{\bm{y}}}
\vphantom{\mbox{\protect\rule{0ex}{1.2em}}} &
_{\renewcommand{\arraystretch}{.7} \begin{array}{c}
\bar{r}_i=\bar{r}_\infty \\
\overline{\gamma}'_{jk}=\overline{\gamma}'_\infty
\end{array} \renewcommand{\arraystretch}{1} } \end{array} \;\;\; \forall \;\; \begin{array}[t]{l} 1 \leq i \leq
N \;\; \mbox{and}
\\
1 \leq j < k \leq N\,. \end{array} \label{eq:ypT}
\end{equation}
Inserting Eq.~(\ref{eq:ytoy}) into Eq.~(\ref{eq:qyt}) one obtains
$ {\bm{q}} = {\bm{q}}_{\infty} + \delta^{1/2} {\bm{q}}^{\prime}$
\,, where ${\bm{q}}_\infty = {\bm{b}}^T {\bar{\bm{y}}}_\infty$\,.
Using these in Eq.~(\ref{eq:wavefunct}) one obtains
\begin{equation} \label{eq:Phi_0q}
\Phi_0({\bar{\bm{y}}}^{\prime})= \hspace{-3ex} \prod_{ \mu = \{
\bm{0}^\pm,\hspace{0.5ex} \bm{1}^\pm,\hspace{0.5ex} \bm{2} \} }
 \,\, \prod_{\xi=1}^{d_{\mu}} \,\, \phi_{n_{\mu_\xi}}
\!\! \left( \left\{D\,\bar{\omega}_\mu\right\}^{1/2}
([{\bm{q}}^\mu]_\xi - {\bm{q}}_\infty^\mu)
 \right) \,.
\end{equation}
Equation~(\ref{eq:Phi_0q}) represents oscillations about the
$D\rightarrow\infty$
configuration ${\bm{q}}_\infty^\mu$ with frequencies
$\{D \, \bar{\omega}_\mu \}$\,. For a macroscopic quantum confined
system at $\delta=1/3$ ($D=3$), $D\,\bar{\omega}_\mu =
3\,\bar{\omega}_\mu$ is sufficiently small that this
harmonic-order, Jacobian-weighted wave function has a macroscopic
extension. However, when $D$ becomes large ($\delta$ becomes
small) the frequencies $\{D \, \bar{\omega}_\mu \}$ becomes very
large and so, according to Eq.~(\ref{eq:Phi_0q}), the harmonic-order
wave function becomes strongly localized about $[{\bm{q}}^\mu]_\xi
= {\bm{q}}_\infty^\mu$ (i.e., it features short wavelengths).

This dimensional behavior
has interesting, and very useful consequences. As is well
known, the energy and density profile of a small-$a$ Bose-Einstein
condensate at $D=3$ depends only on the scattering length of the
interatomic potential, and not the detailed shape of the
potential. This is due to the long wavelength nature of BEC's: for
small to moderate scattering lengths, the atomic wavelengths are
not short enough to ``resolve'' the short-range detail of the
potential. However, for large $D$ the atomic wavelengths become
very short, since according to Eqs.~(\ref{eq:kappascale}) and
(\ref{eq:scale1}), the scaled, Jacobian-transformed Hamiltonian
displays an effective mass term proportional to $D^2$\,. Thus as
we have noted above, and unlike at $D=3$\,, the wave length of the
wave function for the large-$D$ system becomes smaller and becomes
sensitive to the details of the potential.

This feature is actually advantageous. A perturbation theory in
some parameter which at low orders displays an insensitivity to
the precise shape of the interatomic potential could not be
optimized to yield results at low order that would be close to the
actual results. For example, one could not reasonably expect the
energy and density profile at low orders to be both insensitive to
the precise shape of the interatomic potential for small fixed
scattering length and, at the same time, to differ only a small
amount from the actual $D=3$ condensate energy and density
profile. The energy and density profile at low orders would almost
certainly be different from the actual $D=3$ condensate.

The large-$D$ sensitivity to the details of the interatomic
potential in the present method enables us to optimize our
dimensional continuation of the interparticle potential so that
higher-order contributions of this theory will be small.

\section{Application: The Density Profile for an $\bm{N}$-Atom Bose-Einstein
Condensate} \label{sec:BECdensityprofile} We assume a $T=0K$
condensate which is confined by an isotropic, harmonic trap with
frequency $\omega_{ho}$:
\begin{equation}
V_{\mathtt{conf}}(r_i)=\frac{1}{2}m\omega_{ho}^2{r_i}^2\,.
\end{equation}
A realistic two-body atom-atom interaction potential would lead to
a solid-like ground state, and so typically this two-body potential
is replaced with a pseudopotential which does not support bound
states. We follow most other work and replace the actual atom-atom
potential by a hard sphere of radius $a$:
\begin{equation}
V_{\mathtt{int}}(r_{ij})=
\left\{ \begin{array}{ll} \infty\,, & r_{ij} < a \\
0\,,& r_{ij} \ge a\,.
\end{array}\right. \,,
\end{equation}
where $a$ is the s-wave scattering length of the condensate atoms.
We dimensionally continue the hard-sphere potential so that it is
differentiable away from $D=3$\,, allowing us to perform the
dimensional perturbation analysis (see Ref. \cite{FGpaper,energy}
as well as a later discussion in this paper). Thus, we take the
interaction to be
\begin{equation} \label{eq:vint2}
V_{\mathtt{int}}(r_{ij})=\frac{V_o}{1-3/D} \left[1-\tanh \left[
\vphantom{\left(r_{ij}-\alpha-\frac{3}{D}(a-\alpha)\right)}
\frac{c_o}{1-3/D} \left(r_{ij}-a \right) \right] \right] \,,
\end{equation}
%
%
%
where $D$ is the Cartesian dimensionality of space.  This
interaction becomes a hard sphere of radius $a$ in the physical,
$D=3$\,, limit. The two constants, $V_o$ and $c_o$ (which
determine the height and the slope of the potential), are
parameters that allow us to fine-tune the large-$D$ shape of the
potential and optimize our results through harmonic order by
minimizing the contribution of the higher-order terms beyond
harmonic (see Section~\ref{sec:fitting}). Although we have chosen
the simplest possibility for the interatomic potential, two
parameters, we can have any number of parameters providing for a
more general and flexible potential\cite{energy}. The functional
form of the potential at $D\neq 3$ is not unique. Other functional
forms could be chosen with equal success as long as the form is
differentiable and reduces to a hard-sphere potential at $D=3$\,.
We simply choose a form that allows a gradual softening of the
hard wall.

We need to regularize the large-$D$ limit of the Jacobian-weighted
Hamiltonian ($J^{1/2} H J^{-1/2}$). We do this by converting the
variables to dimensionally-scaled harmonic-oscillator units
(bars):
\renewcommand{\arraystretch}{1.5}
\begin{equation} \label{unitsHS}
\begin{array}{l} \bar{r}_i =\frac{r_i}{D^2
\bar{a}_{ho}},\;\;\, \bar{E} = \frac{D^2}{\hbar \bar{\omega}_{ho}}
E, \;\;\,  \bar{H} = \frac{D^2}{\hbar \bar{\omega}_{ho}} H, \;\;\,
\bar{a}
=\frac{a}{\sqrt{2} D^2 \bar{a}_{ho}}, \\
\bar{V}_{o} = \frac{D^2}{\hbar \bar{\omega}_{ho}} V_{o}, \;\;\;
\bar{c}_{o}= \sqrt{2} D^2 \bar{a}_{ho}c_o\,,
\end{array}
\end{equation}
\renewcommand{\arraystretch}{1}
where
\begin{equation} \label{unitsHSa}
\bar{a}_{ho}=\sqrt{\frac{\hbar}{m \bar{\omega}_{ho}}} \;\;\;
\mbox{and} \;\;\; {\bar{\omega}_{ho}}=D^3{\omega_{ho}}
\end{equation}
are the dimensionally-scaled harmonic-oscillator length and
dimensionally-scaled trap frequency, respectively.  In this case
we have chosen $\kappa(D) = D^2 \bar{a}_{\mathtt{ho}}$\,,
while the dimensionally-scaled harmonic-oscillator units of
energy, length and time are $\hbar \bar{\omega}_{ho}$\,,
$\bar{a}_{ho}$\,, and $1/\bar{\omega}_{ho}$\,, respectively.  All
barred constants ($\bar{a}$\,, $\bar{a}_{ho}$\,,
$\bar{\omega}_{ho}$\,, $\bar{V}_o$\,, and $\bar{c}_o$) are held
fixed as $D$ varies. For example, as $D$ varies $\bar{a}$ is held
fixed at a value by requiring that it give the physical unscaled
scattering length at $D=3$\,.

In dimensionally scaled units the total interaction term reads
\begin{eqnarray}
V & = & V_{\mathtt{conf}} + V_{\mathtt{int}} =
\sum\limits_{i=1}^{N}\frac{1}{2}\bar{r}_i^2 +
\frac{\bar{V}_{o}}{1-3\delta}
\sum\limits_{i=1}^{N}\sum\limits_{j=i+1}^{N} \nonumber\\
&& \hspace{2.5em} \left(1-\tanh\left[ \frac{\bar{c}_o}{1-3\delta}
\left( \frac{\bar{r}_{ij}}{\sqrt{2}}-\bar{a} \right)
\vphantom{\sum_{n=1}^{s-3}c_n} \right] \right) \,.
\end{eqnarray}

The infinite-$D$ ($\delta \to 0$) effective potential in
dimensionally-scaled harmonic-oscillator units of
Eqs.~(\ref{unitsHS}) and (\ref{unitsHSa}) is
\begin{eqnarray} \label{veff:HS}
\lefteqn{ \hspace{-5ex}
V_{\mathtt{eff}}=\sum\limits_{i=1}^{N}\left( \frac{1}{8
{\bar{r}_i}^2}\frac{\Gamma^{(i)}}{
\Gamma}+\frac{1}{2}\bar{r}_i^2\right)} \nonumber \\&&
\begin{array}[b]{@{}l@{}l@{}} {\displaystyle \hspace{-4ex} + \bar{V}_{o}
\sum\limits_{i=1}^{N}\sum\limits_{j=i+1}^{N} \left(1-\tanh \left[
\vphantom{\sum_{n=1}^{s-3}c_n} \right. \right. } & {\displaystyle
\left. \left. \bar{c}_o
\left(\frac{\bar{r}_{ij}}{\sqrt{2}}-\bar{a} \right)
\vphantom{\sum_{n=1}^{s-3}c_n} \right] \right) } \end{array} \,.
\end{eqnarray}
One can see from the double-sum term in $V_{\mathtt{eff}}$ that
the large-$D$ interatomic potential has become a soft sphere of
radius  $\bar{a}$ and height $2\bar{V}_{o}$\,. The slope of the
soft wall is determined by $\bar{c}_{o}$\,. The development of DPT
using the three-parameter extension of Eq.~(\ref{eq:vint2}) is
discussed at length in Ref.~\cite{FGpaper}, while a four parameter
extension of Eq.~(\ref{eq:vint2}) is discussed in
Ref.~\cite{energy}.

As noted above, the two parameters in Eq.~(\ref{veff:HS})
are chosen with the goal
of minimizing the contribution of the higher-order beyond-harmonic
terms, and so optimizing the harmonic-order DPT density profile (of
Eq.~(\ref{eq:Nrresult})) and energy perturbation series through
harmonic order in $\delta$ (Eq.~(\ref{eq:E1})).

\section{Optimization of the Interparticle Parameters of a Quantum Confined
System }\label{sec:fitting}
We consider a BEC in a spherical trap
where $\omega_{ho}=2\pi \times 77.87 Hz$ and for which $a=1 \,
000$ a.u. or $0.0433a_{ho}$ in oscillator units
($a_{ho}=\sqrt{\hbar/m\omega_{ho}}$) which is approximately equal
to ten times the natural $^{87}$Rb value. We choose this value for
the scattering length since the actual density profiles and
energies show a noticeable difference from the mean-field, GP
equation result at low $N$\,, and yet the modified
Gross-Pitaevskii (MGP) equation\cite{braaten}
remains valid for comparison to
the DPT result.

The potential is optimized by fitting the energies through
harmonic order to benchmark DMC energies\cite{blume} at low atom number ($N
\le 50$) and DMC density profiles for $N=3$ and $N=10$\,
(DMC density profiles for larger $N$ have not been
obtained\cite{blume}). Since in
our DPT analysis the number of atoms $N$ is a parameter, we can
readily extrapolate to larger $N$ without large amounts of
calculation.

A least-squares fit is used to optimize the parameters of the
dimensionally-continued interatomic potential.  We fit to six
accurate low-$N$ DMC energies\cite{blume} and, as noted above, the position
and height of the peak of the DMC density profile for $N=3$ and
$N=10$\,. The DPT values for the position and height of the peak
of the density profile are just
\renewcommand{\jot}{0.5em}
\begin{eqnarray}
\bar{r}^{(DPT)}_{peak}(N;\bar{V}_o,\bar{c}_o) & = & 
\bar{r}_{\infty}
\hspace{14ex} \label{eq:rDPTpeak} \\
(N_0)^{(DPT)}_{max}(N;\bar{V}_o,\bar{c}_o) && \nonumber \\
\lefteqn{= \left. \frac{N}{S(D)} \, \sqrt{ \frac{\delta \,
\kappa^2(D) \,\, (K_0 - \bm{K}^T \bm{\mathcal{K}}^{-1} \bm{K} )
}{\pi} } \,\, \right|_{D=3} }  \hspace{19ex}
\end{eqnarray}
\renewcommand{\jot}{0em}

Thus we minimize the following quantity with respect to the two
parameters $\bar{V}_o$ and $\bar{c}_0$:
\begin{equation} 
\begin{array}[b]{r@{}c@{}l} {\displaystyle R^2 } & = &  {\displaystyle \sum_{i=1}^6 \left(
\bar{E}^{(DMC)}_{i}-\bar{E}^{(DPT)}(N_i;\bar{V}_o,\bar{c}_o)
\right)^2 }
\\ && {\displaystyle + \hspace{-2ex} \sum_{N=\{3,
10\}} \left[ \left( \bar{r}^{(DMC)}_{peak}(N) -
\bar{r}^{(DPT)}_{peak}(N;\bar{V}_o,\bar{c}_o) \right)^2 \right. }\\
&& {\displaystyle  \left. + \left( N^{(DMC)}_{max}(N) -
(N_0)^{(DPT)}_{max}(N;\bar{V}_o,\bar{c}_o) \right)^2 \right] } \,,
\end{array}
\end{equation}
where $\bar{E}^{(DMC)}_i$ is the dimensionally-scaled DMC energy
for a condensate with atom number $N_i$\,, while
$\bar{r}^{(DMC)}_{peak}(N)$ and $N^{(DMC)}_{max}(N)$ are the
dimensionally-scaled position and height of the peak of the DMC
density profile. The quantity
$\bar{E}^{(DPT)}(N_i;\bar{V}_o,\bar{c}_o)$ is the DPT energy
approximation through harmonic order given by Eqs.~(\ref{eq:E1})
with interatomic potential parameters $\bar{V}_o$ and
$\bar{c}_o$\,, evaluated at $D=3$\,. Thus we have a two-parameter
fit for $\bar{c}_o$ and $\bar{V}_o$\,. This results in values
$c_o=4.7315$ and $V_o=630.9573$ at an $R^2$ of $14.6015$\,.

We remind the reader that the potential at $D = 3$ remains a simple
hard-sphere potential at $r = a$.
\section{Results and Discussion} \label{sec:resultsranddiscussion}
\begin{figure}
\includegraphics[scale=1.0]{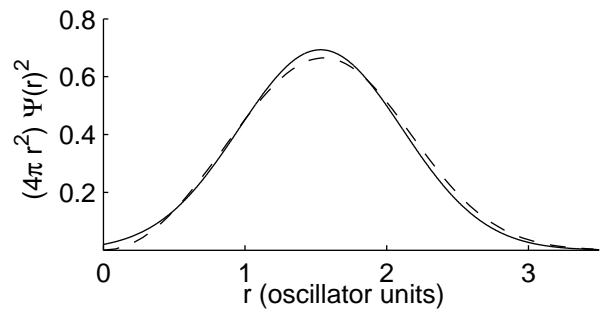}
\renewcommand{\baselinestretch}{.9}
\caption{The harmonic-order number density per atom versus
radial distance of a
spherically confined BEC of 100 $^{87}$Rb atoms with $a=1\,000$
a.u. and $\omega_{ho} = 2\pi \times 77.87$ Hz. The solid line is
the analytic DPT density and the dashed line is the MGP density.
Both curves are weighted by the Jacobian.} \label{fig:one}
\end{figure}
In Fig.~\ref{fig:one} we plot the Jacobian-weighted density
profile divided by the number of atoms $N$ at $N=100$ for a
scattering length $a/a_{ho}=0.0433$, or roughly ten times the
natural scattering length of $^{87}$Rb when $\omega_{ho}=2\pi
\times 77.87 Hz$\,. One hundred atoms is a ten-fold increase over
the largest $N$ density profile initially used to determine the
interatomic potential parameters $\bar{c}_o$ and $\bar{V}_o$\,. As
we see, the DPT density profile lies close to the MGP result,
particularly in the $0.5 \leq r/a_{ho} \leq 1$ range, although the
peak of the DPT profile is a little higher. Since the area under
the curve is normalized to one, the extra height of the DPT peak
comes at the expense of density at larger values of $r/a_{ho}$
which lies below both the GP and MGP result.

\begin{figure}
\includegraphics[scale=1.0]{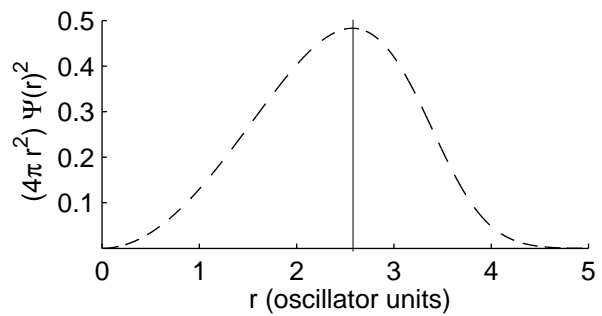}
\renewcommand{\baselinestretch}{.9}
\caption{The MGP number density per atom
versus radial distance of a
spherically confined BEC of 1,000 $^{87}$Rb atoms with $a=1\,000$
a.u. and $\omega_{ho} = 2\pi \times 77.87$ Hz. The dashed line is
the MGP density weighted by the Jacobian. The vertical line is 
inserted to emphasize the asymmetry of the
MGP curve at higher $N$\,.} \label{fig:two}
\end{figure}
At larger values of $N$ the density profile develops an asymmetric
aspect. This is illustrated in Fig.~(\ref{fig:two}) where we plot
the MGP Jacobian-weighted density profile at the same scattering
length for $N=1,000$\,. The lowest-order DPT density profile of
Eq.~(\ref{eq:Nrresult}) is a gaussian and symmetric about
$r^{(DPT)}_{peak}(N;\bar{V}_o,\bar{c}_o)$\,, and so is unable to
capture this emerging asymmetry as $N$ increases. Also if we fix
$N$ and increase the scattering length yet further, the density
profile becomes increasingly asymmetric. Thus while the
harmonic-order DPT density profile is reasonably accurate for
small $a$ and for smaller $N$ at intermediate $a$, incorporating
beyond-mean-field effects for more strongly interacting
systems (larger $N$ or $a$) means we need to go to next order
beyond harmonic in the
DPT perturbation expansion to achieve an asymmetric
beyond-mean-field Jacobian-weighted density profile.

A next-order calculation for more strongly-interacting systems
may be outlined as follows.
The higher-order beyond-harmonic
interaction terms have to be re-expressed in terms of the normal
coordinates of the large-$D$ system. In this regard the $S_N$
symmetry greatly simplifies this task since the interaction terms
individually have to transform under the scalar $[N]$ irreducible
representation of $S_N$\,. In particular, there are only a finite
number of Clebsch-Gordon coefficients of $S_N$ coupling a finite
number of normal coordinates, which transform under the three
above discussed irreducible representations of $S_N$\,, together
to form a scalar $[N]$ irrep. For example, there are only eight
Clebsch-Gordon coefficients of $S_N$ which couple three normal
coordinates transforming under the $[N]$\,, $[N-1, \hspace{1ex}
1]$\,, or $[N-2, \hspace{1ex} 2]$ irreducible representations
together to form a scalar $[N]$ irrep. Actually, it is this limited number
of Clebsch-Gordon coefficients which has allowed for the
essentially analytic solution discussed in this paper of the
harmonic-order DPT approximation and it greatly simplifies the
higher-order beyond-harmonic interaction terms and their necessary
transformation to normal coordinates. The required Clebsch-Gordon
coefficients may be calculated analytically for arbitrary $N$\,.

\section{Conclusions.} \label{sec:Conc}
In this paper we have  calculated an $N$-body, analytic, lowest-order
DPT Jacobian-weighted density profile for a quantum confined system with
a general two-body interaction
from the previously derived analytic lowest-order
DPT wave function\cite{paperI}. The density profile is
directly observable for macroscopic quantum confined systems, such
as a BEC.

The theory applied in this paper for
$L = 0$ states of spherically confined systems is applicable to systems
with attractive or repulsive interparticle interactions
and is also applicable to both
weakly (mean field) and strongly (beyond-mean-field) correlated
systems.
As we have noted above, the fact that
$\overline{\gamma}_\infty$ is not zero is an indication that
beyond-mean-field effects are included in this result even in the
$D\rightarrow\infty$ limit.
This theory is readily
generalizable to systems for which the confining potential has
cylindrical symmetry,

We have applied the general formalism for the harmonic-order DPT
Jacobian-weighted density profile developed in this paper, to the
example of a spherically confined BEC. The higher-order
DPT terms are minimized by analytically
continuing the interatomic potential in $D$ away from a hard
sphere of radius $a$ at $D=3$ so that we fit closely to accurate
low-$N$ DMC energies and Jacobian-weighted density profiles ($N=3$
and $10$ for the density profiles). We have chosen a scattering
length $a/a_{ho}=0.0433$, or roughly ten times the natural
scattering length of $^{87}$Rb when $\omega_{ho}=2\pi \times 77.87
Hz$\,. At this value of $a$ the beyond-mean-field MGP equation
remains valid for comparison. Exploiting the fact that our result
is largely analytic, so that the particle number, $N$\,, is a
parameter in our theory, we have tested the Jacobian-weighted
density profile at $N=100$ and find the DPT result to lie close to
the MGP result.

However, for more strongly interacting systems the Jacobian-weighted density
profile develops an asymmetry about the peak that our
harmonic-order DPT Jacobian-weighted density profile does not
mimic. Thus higher-order calculations are required
for larger $N$ and large $a$\,. A detailed program for calculating
higher-order DPT corrections to $N$-body systems
has been laid out in the paper by Dunn et.\
al.\cite{matrix_method} and has been applied to high order for
small-$N$ systems\cite{highorder}. As discussed at the end of
Section~\ref{sec:resultsranddiscussion} of this paper, for
large-$N$ systems the $S_N$ point-group symmetry is at the heart
of, and greatly simplifies, the calculation of these higher-order
terms.

\section{Acknowledgments}

We acknowledge continued support from the Army Research Office.  We thank
Doerte Blume for DMC results.


\begin{thebibliography}{99}
\bibitem{footn} Actually, in the case of CH$_4$ the carbon atom should be
positioned away from the four hydrogens in a fourth dimension
orthogonal the three-dimensional subspace spanned by the hydogen
atoms. The position of the carbon in the actual $D=3$ methane
molecule represent the point of intersection of this fourth
dimension in which the carbon atom sits with the three-dimensional
subspace spanned by the hydogen atoms.
\end{thebibliography}
\end{document}